\documentclass[a4paper,reqno]{amsart}
\usepackage{amsart-color}      
\usepackage[all]{xy}           
\usepackage{amssymb}           
\usepackage{hyperref}
\usepackage{eucal}

\numberwithin{equation}{section}

\newtheorem{definition}{Definition}[section]

\newtheorem{theorem}[definition]{Theorem}
\newtheorem{proposition}[definition]{Proposition}
\newtheorem{corollary}[definition]{Corollary}
\newtheorem{remarkth}[definition]{Remark}

\newenvironment{remark}{\begin{remarkth}\upshape}{\hfill$\diamond$\end{remarkth}}

\renewcommand{\emph}[1]{{\bfseries\itshape{#1}}}

\newcommand{\R}{\mathbb{R}}      
\newcommand{\F}{\mathbb{F}}

\newcommand{\lcf}{\lbrack\! \lbrack}
\newcommand{\rcf}{\rbrack\! \rbrack}
\let\maparrow\longrightarrow
\let\to\maparrow
\newcommand\map[3]{#1 \colon #2\maparrow#3}
\newcommand{\lvec}[1]{\overleftarrow{#1}}
\newcommand{\rvec}[1]{\overrightarrow{#1}}

\renewcommand{\d}{\mathrm{d}^\circ}
\newcommand{\qquand}{\qquad\text{and}\qquad}
\newcommand{\quand}{\quad\text{and}\quad}

\newcommand{\tr}{\operatorname{Tr}}

\makeatletter
\newcommand\prol{\@ifstar{\@proldf}{\@prolpf}}  
\def\@prolpf{\@ifnextchar[{\@prolpf@wrt}{\@prolpf@}}
\def\@prolpf@wrt[#1]#2{\@ifnextchar[{\@prolpf@wrt@at{#1}{#2}}{\@prolpf@wrt@{#1}{#2}}}
\def\@prolpf@wrt@at#1#2[#3]{\prolsymbol^{#1}_{#3}#2}
\def\@prolpf@wrt@#1#2{\prolsymbol^{#1}#2}
\def\@prolpf@#1{\@ifnextchar[{\@prolpf@at{#1}}{\@prolpf@@{#1}}}
\def\@prolpf@at#1[#2]{\prolsymbol_{#2}#1}
\def\@prolpf@@#1{\prolsymbol#1}
\def\@proldf{\@ifnextchar[{\@proldf@wrt}{\@proldf@}}
\def\@proldf@wrt[#1]#2{\@ifnextchar[{\@proldf@wrt@at{#1}{#2}}{\@proldf@wrt@{#1}{#2}}}
\def\@proldf@wrt@at#1#2[#3]{\prolsymbol^{*#1}_{#3}#2}
\def\@proldf@wrt@#1#2{\prolsymbol^{*#1}#2}
\def\@proldf@#1{\@ifnextchar[{\@proldf@at{#1}}{\@proldf@@{#1}}}
\def\@proldf@at#1[#2]{\prolsymbol^*_{#2}#1}
\def\@proldf@@#1{\prolsymbol^*#1}
\def\prolsymbol{\mathcal{T}}
\makeatother
\newcommand{\TEE}[1][]{\prolsymbol^E_{#1}E}

\newcommand{\X}{\mathcal{X}}
\newcommand{\V}{\mathcal{V}}

\newcommand{\Gc}{\mathcal{G}}
\newcommand{\cnabla}{\check{\nabla}}

\newcommand{\D}{\mathcal{D}} 

\newcommand{\cm}{\mathcal{M}} 
\newcommand{\vd}[1][]{\mathcal{V}_{#1}} 
\newcommand{\cf}{\boldsymbol{\Psi}} 
\def\lcf{\lbrack\! \lbrack}
\def\rcf{\rbrack\! \rbrack}
\newcommand{\spn}{\operatorname{span}}

\newcommand{\Pc}{{\mathcal{P}}}

\def\cnabla{\check{\nabla}}
\newcommand{\pai}[2]{\langle #1, #2\rangle}
\newcommand{\setdef}[2]{\{#1 \; | \; #2\}}

\newcommand{\cinfty}[1]{C^\infty(#1)}
\newcommand{\set}[2]{\left\{\,#1\left.\vphantom{#1#2}\,\right\vert\,#2\,
                \right\}}
\newcommand{\real}{\mathbb{R}}

\renewcommand{\natural}{\mathbb{N}}

\newcommand{\invclos}[1]{\operatorname{\overline{Lie}}(#1)}
\newcommand{\symclos}[1]{\operatorname{\overline{Sym}}(#1)}

\newcommand{\grad}{\operatorname{grad}}
\newcommand{\psinv}{\operatorname{pseudoinv}}

\newcommand{\symprod}[2]{\langle #1: #2\rangle}
\newcommand{\pd}[2]{\frac{\partial #1}{\partial #2}}
\newcommand{\at}[1]{\Big|_{#1}}
\newcommand{\vectorfields}[1]{\mathfrak{X}(#1)}
\newcommand{\Sec}[1]{\operatorname{Sec}(#1)}
\let\sec\Sec

\newcommand{\Hor}{\operatorname{Hor}}
\newcommand{\Ver}{\operatorname{Ver}}
\newcommand{\hor}{\operatorname{hor}}
\newcommand{\ver}{\operatorname{ver}}
\newcommand{\Br}{\operatorname{Br}}
\newcommand{\famX}{\mathcal{X}}
\newcommand{\Id}[1]{\operatorname{id}_{#1}}
\newcommand{\sode}{{\textsc{sode}}}
\newcommand{\spV}{^{\scriptscriptstyle V}}
\newcommand{\spC}{^{\scriptscriptstyle C}}

\newcommand{\CC}{\Gamma} 
\newcommand{\SC}{f} 

\renewcommand{\SC}{S} 
\newcommand{\id}{{\operatorname{id}}}
\newcommand{\dpd}[2]{{\displaystyle\pd{#1}{#2}}}
\def\p#1{\mathaccent19{#1}}
\newcommand{\pb}{^\star} 
\newcommand{\ext}[2][]{\bigwedge\nolimits^{#1}{#2}}
\newcommand{\XsigmaJ}{{X_\sigma^{\scriptscriptstyle(1)}}}
\newcommand{\Real}{\mathbb{R}}
\newcommand{\Lprol}[1]{\mathcal{L}#1}
\newcommand{\Jprol}[1]{\mathcal{J}#1}
\newcommand{\prEM}{\tau^E_M}
\newcommand{\prFN}{\tau^F_N}
\newcommand{\prEF}{\pi}
\newcommand{\prMN}{\nu}
\newcommand{\Lpi}[1][]{\mathcal{L}_{#1}\prEF}
\newcommand{\Jpi}[1][]{\mathcal{J}_{#1}\prEF}
\newcommand{\Vpi}[1][]{\mathcal{V}_{#1}\prEF}

\newcommand{\prLM}{\tilde{\prEF}_{10}}
\newcommand{\prVM}{\boldsymbol{\prEF}_{10}}
\newcommand{\prJM}{\prEF_{10}}
\newcommand{\prJN}{\prEF_1}
\newcommand{\dLpi}[1][]{\mathcal{L}_{#1}^*\!\prEF}

\newcommand{\Mor}[1]{\mathcal{M}(#1)}

\newcommand{\Gprol}[2][G]{#2\overset{#1}{\times}#2}

\def\prolsymbol{\CMcal{T}}

\begin{document}

\title[A survey of Mechanics on Lie algebroids and groupoids]{A survey
  of Lagrangian Mechanics and control on Lie algebroids and groupoids}

\author[J. Cort\'es]{Jorge Cort\'es} \address{Jorge Cort\'es: Applied
  Mathematics and Statistics, Baskin School of Engineering, University
  of California, Santa Cruz, California 95064, USA}
\email{jcortes@ucsc.edu}

\author[M. de Le\'on]{Manuel de Le\'on} \address{M. de Le\'on:
  Instituto de Matem\'aticas y F{\'\i}sica Fundamental, Consejo
  Superior de Investigaciones Cient\'{\i}ficas, Serrano 123, 28006
  Madrid, Spain} \email{mdeleon@imaff.cfmac.csic.es}

\author[J.\ C.\ Marrero]{Juan C.\ Marrero} \address{Juan C.\ Marrero:
  Departamento de Matem\'atica Fundamental, Facultad de
  Ma\-te\-m\'a\-ti\-cas, Universidad de la Laguna, La Laguna,
  Tenerife, Canary Islands, Spain} \email{jcmarrer@ull.es}

\author[D.\ Mart\'{\i}n de Diego]{D. Mart\'{\i}n de Diego}
\address{D.\ Mart\'{\i}n de Diego: Instituto de Matem\'aticas y
  F{\'\i}sica Fundamental, Consejo Superior de Investigaciones
  Cient\'{\i}ficas, Serrano 123, 28006 Madrid, Spain}
\email{d.martin@imaff.cfmac.csic.es}

\author[E.\ Mart\'{\i}nez]{Eduardo Mart\'{\i}nez} \address{Eduardo
  Mart\'{\i}nez: Departamento de Matem\'atica Aplicada, Facultad de
  Ciencias, Universidad de Zaragoza, 50009 Zaragoza, Spain}
\email{emf@unizar.es}


\thanks{This work has been partially supported by
  MEC (Spain) Grants MTM 2004-7832, BFM2003-01319 and BFM2003-02532}

\begin{abstract}
  In this survey, we present a geometric description of Lagrangian and
  Hamiltonian Mechanics on Lie algebroids.  The flexibility of the Lie
  algebroid formalism allows us to analyze systems subject to
  nonholonomic constraints, mechanical control systems, Discrete
  Mechanics and extensions to Classical Field Theory within a single
  framework.  Various examples along the discussion illustrate the
  soundness of the approach.
\end{abstract}

\maketitle

\vspace{-20pt}

\tableofcontents

\section{Introduction}
The theory of Lie algebroids and Lie groupoids has proved to be very
useful in different areas of mathematics including algebraic and
differential geometry, algebraic topology, and symmetry analysis. In
this survey, we illustrate the wide range of applications of this
formalism to Mechanics.  Specifically, we show how the flexibility
provided by Lie algebroids and groupoids allows us to analyze, within
a single framework, different classes of situations such as systems
subject to nonholonomic constraints, mechanical control systems,
Discrete Mechanics and Field Theory.

The notions of Lie algebroid and Lie groupoid allow to study general
Lagrangian and Hamiltonian systems beyond the ones defined on the
tangent and cotangent bundles of the configuration manifold,
respectively. These include systems determined by Lagrangian and
Hamiltonian functions defined on Lie algebras, Lie groups, Cartesian
products of manifolds, and reduced spaces.

The inclusive feature of the Lie algebroid formalism is particularly
relevant for the class of Lagrangian systems invariant under the
action of a Lie group of symmetries.  Given a standard Lagrangian
system, one associates to the Lagrangian function a Poincar\'e-Cartan
symplectic form and an energy function using the particular geometry
of the tangent bundle.  The dynamics is then obtained as the
Hamiltonian vector field associated to the energy function through the
Poincar\'e-Cartan form.  The reduction by the Lie group action of the
dynamics of this system yields a reduced dynamics evolving on a
quotient space (which is not a tangent bundle). However, the interplay
between the geometry of this quotient space and the reduced dynamics
is not as transparent as in the tangent bundle case.  Recent efforts
have lead to a unifying geometric framework to overcome this drawback.
It is precisely the underlying structure of Lie algebroid on the phase
space what allows a unified treatment. This idea was introduced by
Weinstein~\cite{weinstein} (see also~\cite{L}), who developed a
generalized theory of Lagrangian Mechanics on Lie algebroids. He
obtained the equations of motion using the linear Poisson structure on
the dual of the Lie algebroid and the Legendre transformation
associated with a (regular) Lagrangian. In~\cite{weinstein}, Weinstein
also posed the question of whether it was possible to develop a
treatment on Lie algebroids and groupoids similar to Klein's formalism
for ordinary Lagrangian Mechanics~\cite{Klein}.  This question was
answered positively by E.  Mart{\'\i}nez in~\cite{M} (see
also~\cite{CoMa,Ma2,Ma4,PP}). The main notion was that of prolongation
of a Lie algebroid over a mapping, introduced by P.J.  Higgins and K.
Mackenzie~\cite{HM}.  More recently, the work~\cite{LeMaMa} has
developed a description of Hamiltonian and Lagrangian dynamics on a
Lie algebroid in terms of Lagrangian submanifolds of symplectic Lie
algebroids. An alternative approach, using the linear Poisson
structure on the dual of the Lie algebroid was discussed
in~\cite{Grabo}.

In this paper, we present an overview of these developments. We make
special emphasis on one of the main advantages of the Lie algebroid
formalism: the possibility of establishing appropriate maps (called
morphisms) between systems that respect the structure of the phase
space, and allow to relate their respective properties. As we will
show, this will allow us to present a comprehensive study of the
reduction process of Lagrangian systems while remaining within the
same category of mathematical objects.

We also consider nonholonomic systems (i.e., systems subject to
constraints involving the velocities, see~\cite{CoLeMaMa} for a list
of references) and control systems evolving on Lie
algebroids~\cite{CoMa}. This is motivated by the renewed interest in
the study of nonholonomic mechanical systems for new applications in
the areas of robotics and control.  In particular, we provide widely
applicable tests to decide the accessibility and controllability
properties of mechanical control systems defined on Lie algebroids.

We end this survey by paying attention to two recent developments in
the context of Lie algebroids and groupoids: Discrete Mechanics and
Classical Field Theory.  Discrete Mechanics seeks to develop a
complete discrete-time counterpart of the usual continuous-time
treatment of Mechanics. The ultimate objective of this effort is the
construction of numerical integrators for Lagrangian and Hamiltonian
systems (see~\cite{mawest} and references therein). Up to now, this
effort has been mainly focused on the case of discrete Lagrangian
functions defined on the Cartesian product of the configuration
manifold with itself.  This Cartesian product is just an example of a
Lie groupoid.  Here, we review the recent developments
in~\cite{groupoid}, where we proposed a complete description of
Lagrangian and Hamiltonian Mechanics on Lie groupoids.  In particular,
this description covers the analysis of discrete systems with
symmetries, and naturally produces reduced geometric integrators.
Another extension that we consider is the study of Classical Field
Theory on Lie algebroids.  Thinking of a Lie algebroid as a substitute
of the tangent bundle of a manifold, we substitute the classical
notion of fibration bundle by a surjective morphism of Lie algebroids
$\map{\pi}{E}{F}$. Then, we construct the jet space $\Jpi$ as the
affine bundle whose elements are linear maps from a fiber of $F$ to a
fiber of $E$, i.e., sections of the projection $\pi$.  After a
suitable choice of the space of variations, we derive the
Euler-Lagrange equations for this problem.

The paper is organized as follows.  In Section~\ref{preliminaries} we
present some basic facts on Lie algebroids, including results from
differential calculus, morphisms and prolongations of Lie algebroids,
and linear connections. We also introduce the notion of Lie groupoid,
Lie algebroid associated to a Lie groupoid, and morphisms and
prolongations of Lie groupoids.  Various examples are given in order
to illustrate the generality of the theory.  In
Section~\ref{Mechanics}, we give a brief introduction to the
Lagrangian formalism of Mechanics on Lie algebroids, determined by a
Lagrangian function $L : E \to \R$ on the Lie algebroid $\tau : E \to
M$.  Likewise, we introduce the Hamiltonian formalism on Lie
algebroids, determined by a Hamiltonian function $\map{H}{E^*}{\R}$,
where $\map{\tau^*}{E^*}{M}$ is the dual of the Lie algebroid $E \to
M$.  In Section~\ref{nonlinear} we introduce the class of nonholonomic
Lagrangian systems.  We study the existence and uniqueness of
solutions, and characterize the notion of regularity of a nonholonomic
system. Under this property, we derive a procedure to obtain the
solution of the nonholonomic problem from the solution of the free
problem by means of projection techniques. Moreover, we construct a
nonholonomic bracket that measures the evolution of the observables,
and we study the reduction of nonholonomic systems in terms of
morphisms of Lie algebroids.  In Section~\ref{control} we introduce
the class of mechanical control systems defined on a Lie algebroid. We
generalize the notion of affine connection control system to the
setting of Lie algebroids, and introduce the notions of (base)
accessibility and (base) controllability. We provide sufficient
conditions to check these properties for a given mechanical control
system.  In Section~\ref{discrete}, we study Discrete Mechanics on Lie
groupoids.  In particular, we construct the discrete Euler-Lagrange
equations, discrete Poincar\'e-Cartan sections, discrete Legendre
transformations, and Noether's theorem, and identify the preservation
properties of the discrete flow. In the last section, we extend the
variational formalism for Classical Field Theory to the setting of Lie
algebroids. Given a Lagrangian function, we study the problem of
finding critical points of the action functional when we restrict the
fields to be morphisms of Lie algebroids. Throughout the paper,
various examples illustrate the results. We conclude the paper by
identifying future directions of research.

\section{Lie algebroids and Lie groupoids}\label{preliminaries}

\subsection{Lie algebroids}

Given a real vector bundle $\map{\tau}{E}{M}$, let $\Sec{\tau}$ denote
the space of the global cross sections of $\map{\tau}{E}{M}$.  A
\emph{Lie algebroid} $E$ over a manifold $M$ is a real vector bundle
$\map{\tau}{E}{M}$ together with a Lie bracket $\lcf\cdot, \cdot\rcf$
on $\Sec{\tau}$ and a bundle map $\map{\rho}{E}{TM}$ over the
identity, called \emph{the anchor map}, such that the homomorphism
(denoted also $\map{\rho}{\Sec{\tau}}{\mathfrak{X}(M)}$) of
$C^\infty(M)$-modules induced by the anchor map verifies
\[
\lcf X,fY\rcf=f\lcf X,Y\rcf + \rho(X)(f)Y,
\]
for $X,Y\in \Sec{\tau}$ and $f\in C^\infty(M)$. The triple
$(E,\lcf\cdot,\cdot\rcf,\rho)$ is called \emph{a Lie algebroid
over} $M$ (see \cite{Ma,Mac}). If $(E,\lcf\cdot,\cdot\rcf,\rho)$
is a Lie algebroid over $M,$ then the anchor map
$\map{\rho}{\Sec{\tau}}{\mathfrak{X}(M)}$ is a homomorphism
between the Lie algebras $(\Sec{\tau},\lcf\cdot,\cdot\rcf)$ and
$(\mathfrak{X}(M),[\cdot,\cdot])$.

In what concerns to Mechanics, it is convenient to think of a Lie
algebroid as a generalization of the tangent bundle of $M$. One
regards an element $a$ of $E$ as a generalized velocity, and the
actual velocity $v$ is obtained when applying the anchor to $a$, i.e.,
$v=\rho(a)$. A curve $\map{a}{[t_0,t_1]}{E}$ is said to be
\emph{admissible} if $\dot{m}(t)=\rho(a(t))$, where $m(t)=\tau(a(t))$
is the base curve.

Given local coordinates $(x^i)$ in the base manifold $M$ and a local
basis of sections $(e_{\alpha})$ of $E$, then local coordinates of a
point $a\in E$ are $(x^i, y^{\alpha})$ where $a=y^{\alpha}
e_{\alpha}(\tau(a))$. In local form, the Lie algebroid structure is
determined by the local functions $\rho _{\alpha}^{i}$ and
$C_{\alpha\beta}^{\gamma}$ on $M$. Both are determined by the
relations
\begin{eqnarray}
  \rho (e_{\alpha}) &=&\rho _{\alpha}^{i}\frac{\partial}{\partial x^i}
  ,  \label{anch} \\
  \lcf e_{\alpha}, e_{\beta}\rcf
  &=&C_{\alpha\beta}^{\gamma}\ e_{\gamma}
  \label{liea}
\end{eqnarray}
and they satisfy the following equations
\begin{equation}
  \rho _{\alpha}^{j}\frac{\partial \rho _{\beta}^{i}}{\partial
    x^{j}}-\rho _{\beta}^{j}%
  \frac{\partial \rho _{\alpha}^{i}}{\partial x^{j}}=\rho
  _{\gamma}^{i}C_{\alpha\beta}^{\gamma}~%
    \mbox{\ and\
  }\sum\limits_{\hbox{cyclic}(\alpha,\beta,\gamma)}\left[
    \rho _{\alpha}^{i}\frac{\partial C_{\beta
        \gamma}^{\nu}}{\partial
      x^{i}}+C_{\beta\gamma}^{\mu}C_{\alpha\mu}^{\nu}\right]
  =0. \label{lasa}
\end{equation}

\paragraph{\textbf{Cartan calculus}} One may define
\emph{the exterior differential of $E$}, $\map{d}{\Sec{\wedge^k
    \tau^*}}{\Sec{\wedge^{k+1}\tau^*}}$, as follows
\begin{equation}\label{dif}
\begin{aligned}
d \omega(X_0,\dots, X_k)&=\sum_{i=0}^{k}
(-1)^i\rho(X_i)(\omega(X_0,\dots,
\widehat{X_i},\dots, X_k)) \\
&+ \sum_{i<j}(-1)^{i+j}\omega(\lcf
X_i,X_j\rcf,X_0,\dots,
\widehat{X_i},\dots,\widehat{X_j},\dots ,X_k),
\end{aligned}
\end{equation}
for $\omega\in \Sec{\wedge^k \tau^*}$ and $X_0,\dots ,X_k\in
\Sec{\tau}.$ $d$ is a cohomology operator, that is, $d^2=0$. In
particular, if $f: M\to \R$ is a real smooth function then
$df(X)=\rho(X)f,$ for $X\in \Sec{\tau}$. Locally,
\[
dx^i=\rho^i_{\alpha}e^{\alpha}\qquand de^{\gamma}=-\frac{1}{2}
C^{\gamma}_{\alpha\beta} e^{\alpha}\wedge e^{\beta},
\]
where $\{e^{\alpha}\}$ is the dual basis of $\{e_{\alpha}\}$.  We may
also define the Lie derivative with respect to a section $X$ of $E$ as
the operator $\map{\mathcal{L}_X}{\Sec{\wedge^k \tau^*}}{\Sec{\wedge^k
    \tau^*}}$ given by ${\mathcal L}_X=i_X\circ d+d\circ i_X$ (for
more details, see \cite{Ma,Mac}).

\paragraph{\textbf{Morphisms}} Let $(E,\lcf\cdot,\cdot \rcf,\rho)$ (resp.,
$(E',\lcf\cdot,\cdot\rcf', \rho')$) be a Lie algebroid over a manifold
$M$ (resp., $M'$) and suppose that $\map{\Psi}{E}{E'}$ is a vector
bundle morphism over the map $\map{\Psi_{0}}{M}{M'}$. Then, the pair
$(\Psi, \Psi_{0})$ is said to be a \emph{Lie algebroid morphism} if
\begin{equation}\label{dd'}
d ((\Psi, \Psi_{0})^*\phi')= (\Psi,
\Psi_{0})^*(d'\phi'), \;\;\; \text{ for all
}\phi'\in \Sec{\wedge^k(E')^*} \text{ and
for all }k,
\end{equation}
where $d$ (resp., $d'$) is the differential of the Lie algebroid
$E$ (resp., $E'$) (see \cite{LeMaMa}). Note that $(\Psi,
\Psi_{0})^*\phi'$ is the section of the vector bundle $\wedge^k
E^*\maparrow M$ defined for $k>0$ by
\[
((\Psi, \Psi_{0})^*\phi')_x(a_1, \dots
,a_k)=\phi'_{\Psi_{0}(x)}(\Psi(a_1), \dots , \Psi(a_k)),
\]
for $x\in M$ and $a_1,\dots ,a_k\in E_{x}$, and by
$(\Psi,\Psi_0)^*f=f\circ\Psi_0$ for $f\in\Sec{\wedge^0
  E'^*}=\cinfty{M'}$. In the particular case when $M = M'$ and
$\Psi_{0} = id_M$ then (\ref{dd'}) holds if and only if
\[
\lcf \Psi \circ X, \Psi \circ Y \rcf' = \Psi \lcf
X, Y \rcf, \makebox[.3cm]{} \rho'(\Psi X) =
\rho(X), \makebox[.3cm]{} \mbox{ for } X, Y \in
\Sec{\tau}.
\]

\paragraph{\textbf{Linear connections on Lie algebroids}} Let $\tau: E\to M$
be a Lie algebroid over $M$. A \emph{connection on} $E$ is a
$\R$-bilinear map $\map{\nabla}{\Sec{E}\times\Sec{E}}{\Sec{E}}$ such that
\[
\nabla_{fX} Y=f\nabla_X Y\;,\qquad \nabla_X(fY)=\rho(X)(f) Y+f\nabla_X
Y
\]
for $f\in C^{\infty}(M)$ and $X, Y\in\Sec{E}$.

Given a local basis $\{e_{\alpha}\}$ of $\Sec{E}$ such
that $X=X^{\alpha} e_{\alpha}$ and $Y=Y^{\beta} e_{\beta}$ then
\[
\nabla_X Y= X^{\alpha}\left(\rho^i_{\alpha}\frac{\partial
Y^{\gamma}}{\partial x^i}+\Gamma^{\gamma}_{\alpha\beta}
Y^{\beta}\right)e_{\gamma}.
\]
The terms $\Gamma^{\gamma}_{\alpha\beta}$ are called the
\emph{connection coefficients}.  The \emph{symmetric product}
associated with $\nabla$ is given by
\[
\langle X: Y\rangle =\nabla_X Y+ \nabla_Y X,\qquad X, Y\in
\Sec{E}.
\]

Since the connection is $C^{\infty}(M)$-linear in the first argument,
it is possible to define the derivative of a section $Y\in \Sec{E}$
with respect to an element $a\in E_m$ by simply putting
\[
\nabla_a Y=(\nabla_X Y)(m),
\]
with $X\in \Sec{E}$ satisfying $X(m)=a$. Moreover, the connection
allows us to take the derivative of sections along maps and, as a
particular case, of sections along curves. If we have a morphism of
Lie algebroids $\Phi: F\to E$ over the map $\varphi: N\to M$ and a
section $X: N\to E$ along $\varphi$ (i.e., $X(n)\in E_{\varphi(n)}$,
for $n\in N$), then $X$ may be written as
\[
X=\sum_{l=1}^p F_l (X_l\circ \varphi) ,
\]
for some sections $\{X_1, \ldots, X_p\}$ of $E$ and for some
functions $F_1, \ldots, F_p\in C^{\infty}(N)$, and the derivative
of $X$ along $\varphi$ is given by
\[
\nabla_b X=\sum_{l=1}^p\left[ (\rho_F(b)F_l) X_l
(\varphi(n))+F_l(n)\nabla_{\Phi(b)} X_l\right], \quad \hbox{for }
b\in F_n\; ,
\]
where $\rho_F$ is the anchor map of the Lie algebroid $F\to N$.

A particular case of the above general situation is the following.
Let $a: I\to E$ be an admissible curve and $b: I\to E$ be a curve
in $E$, both of them projecting by $\tau$ onto the same base curve
in $M$, $\tau(a(t))=m(t)=\tau(b(t))$. Take the Lie algebroid
structure $TI\to I$ and consider the morphism $\Phi: TI\to E$,
$\Phi(t,\dot{t})=\dot{t} a(t)$ over $m: I\to M$.  Then one can
define the derivative of $b(t)$ along $a(t)$ as $\nabla_{d/dt}
b(t)$. This derivative is usually denoted by $\nabla_{a(t)}b(t)$.
In local coordinates, this reads
\[
\nabla_{a(t)}b(t)=\left[ \frac{d
    b^{\gamma}}{dt}+\Gamma^{\gamma}_{\alpha\beta}
  a^{\alpha}b^{\beta}\right] e_{\gamma}(m(t)), \hbox{ for all } t.
\]
The admissible curve $a: I\to E$ is said to be a \emph{geodesic}
for $\nabla$ if $\nabla_{a(t)}a(t)=0$ (see~\cite{CoMa}).

Now, let ${\mathcal G}: E\times_M E\to \R$ be a bundle metric on a Lie
algebroid $\tau: E\to M$. In a parallel way to the situation in the
tangent bundle geometry, one can see that there is a canonical
connection $\nabla^\Gc$ on $E$ associated with $\Gc$. In fact, the
connection $\nabla^\Gc$ is determined by the formula
\begin{eqnarray*}
  2\Gc(\nabla^\Gc_X Y, Z)&=& \rho(X)(\Gc(Y, Z))+\rho(Y)(\Gc(X,
  Z))-\rho(Z)(\Gc(X,
  Y))\\
  && +\Gc(X, \lcf Z,Y\rcf)+  \Gc(Y, \lcf Z,X\rcf) -\Gc(Z, \lcf
  Y,X\rcf) ,
\end{eqnarray*}
for $X, Y, Z\in\Sec{E}$. $\nabla^\Gc$ is a torsion-less connection
and it is metric  with respect to $\Gc$. In other words
\begin{eqnarray*}
\lcf X, Y\rcf&=&\nabla^\Gc_X Y - \nabla^\Gc_Y X\; ,\\
\rho(X) (\Gc(Y, Z))&=&\Gc (\nabla^\Gc_X Y, Z)+ \Gc (Y,
\nabla^\Gc_X Z)\, .
\end{eqnarray*}
$\nabla^\Gc$ is called the \emph{Levi-Civita connection} of $\Gc$
(see \cite{CoMa}).

Finally, suppose that  $E=D\oplus D^c$, with $D$ and $D^c$ vector
subbundles of $E$, and denote by $P: E\to D$ and $Q: E\to D^c$ the
corresponding complementary projectors induced by the
decomposition. Then, the \emph{constrained connection} is the
connection $\cnabla$ on $E$ defined by
\[
\cnabla_X Y=P(\nabla_X Y)+\nabla_X(QY) ,
\]
for $X, Y\in \Sec{E}$ (for the properties of the constrained
connection $\cnabla$, see \cite{CoMa}).

\paragraph{\textbf{Examples}}
We will present some examples of Lie algebroids.

1.- {\bf Real Lie algebras of finite dimension}. Let $\mathfrak{g}$ be
a real Lie algebra of finite dimension. Then, it is clear that
$\mathfrak{g}$ is a Lie algebroid over a single point.

2.- {\bf The tangent bundle}. Let $TM$ be the tangent bundle of a
manifold $M$. Then, the triple $(TM, [\cdot , \cdot], id_{TM})$ is
a Lie algebroid over $M$, where $id_{TM}: TM \to TM$ is the
identity map.

3.- {\bf Foliations}. Let ${\mathcal F}$ be a foliation of finite
dimension on a manifold $P$ and $\tau_{\mathcal F}: T{\mathcal F} \to
P$ be the tangent bundle to the foliation ${\mathcal F}$. Then,
$\tau_{\mathcal F}: T{\mathcal F} \to P$ is a Lie algebroid over $P$.
The anchor map is the canonical inclusion $\rho_{\mathcal F}:
T{\mathcal F} \to TP$ and the Lie bracket on the space
$\Sec{\tau_{\mathcal F}}$ is the restriction to $\Sec{\tau_{\mathcal
    F}}$ of the standard Lie bracket of vector fields on $P$.  In
particular, if $\pi: P \to M$ is a fibration, $\tau_{P}: TP \to P$ is
the canonical projection and $(\tau_{P})_{|V\pi}: V\pi \to P$ is the
restriction of $\tau_{P}$ to the vertical bundle to $\pi$, then
$(\tau_{P})_{|V\pi}: V\pi \to P$ is a Lie algebroid over~$P$.

4.- {\bf Atiyah algebroids}. Let $p:Q\to M$ be a principal
$G$-bundle. Denote by $\Phi:G\times Q\to Q$ the free action of $G$
on $Q$ and by $T\Phi:G\times TQ\to TQ$ the tangent action of $G$
on $TQ.$ Then, one may consider the quotient vector bundle
$\tau_{Q}|G:TQ/G\to M=Q/G$, and the sections of this vector bundle
may be identified with the vector fields on $Q$ which are
invariant under the action $\Phi$. Using that every $G$-invariant
vector field on $Q$ is $p$-projectable and that the usual Lie
bracket on vector fields is closed with respect to $G$-invariant
vector fields, we can induce a Lie algebroid structure on $TQ/G$.
This Lie algebroid is called \emph{the Atiyah algebroid associated
with the principal $G$-bundle $p:Q\to M$} (see \cite{LeMaMa,Ma}).

5.- {\bf Action Lie algebroids}. Let $(E,\lcf\cdot,\cdot\rcf,\rho)$ be
a Lie algebroid over a manifold $M$ and $\map{f}{M'}{M}$ be a smooth
map. Then, the pull-back of $E$ over $f$, $f^*E=\set{(x',a)\in
  M'\times E}{f(x')=\tau(a)},$ is a vector bundle over $M'$ whose
vector bundle projection is the restriction to $f^*E$ of the first
canonical projection $pr_1:M'\times E\rightarrow M'$. However, $f^*E$
is not, in general, a Lie algebroid.

Now, suppose that $\map{\Phi}{\Sec{\tau}}{\mathfrak{X}(M')}$ is an
action of $E$ on $f$, that is, $\Phi$ is a $\R$-linear map which
satisfies the following conditions
$$\Phi(hX)=(h\circ f)\Phi X,\;\;\Phi\lcf X,Y\rcf=[\Phi X,\Phi
Y],\;\;\Phi X(h\circ f)=\rho(X)(h)\circ f,$$
for $X,Y\in\Sec{\tau}$
and $h\in C^{\infty}(M)$. Then, one may introduce a Lie algebroid
structure $(\lcf\cdot,\cdot\rcf_{\Phi},\rho_{\Phi})$ on the vector
bundle $f^*E\rightarrow M'$ which is characterized by the following
conditions
\begin{equation}\label{2.3'}\lcf X\circ f,Y\circ
f\rcf_{\Phi}=\lcf X,Y\rcf\circ
f,\;\;\rho_{\Phi}(X\circ
f)=\Phi(X),\makebox[1cm]{for} X,Y\in{\rm
Sec}(\tau).
\end{equation}

The resultant Lie algebroid is denoted by $E\ltimes f$ and we call it
\emph{an action Lie algebroid} (for more details, see \cite{LeMaMa}).

6.- {\bf The prolongation of a Lie algebroid over a fibration}
\cite{HM,LeMaMa,Ma2}. Let $(E, \lcf\cdot, \cdot\rcf, \rho)$ be a Lie
algebroid over a manifold $M$ and $\pi: P \to M$ be a fibration. We
consider the subset of $E \times TP$
\[
\prol[E]{P}[p] =\set{(b,v)\in E_x\times T_pP}{\rho(b)=T_p\pi(v)},
\]
where $T\pi: TP \to TM$ is the tangent map to $\pi$, $p\in P_x$ and
$\pi(p)=x$ . We will frequently use the redundant notation $(p,b,v)$
to denote the element $(b,v)\in\prol[E]{P}[p]$.
$\prol[E]{P}=\cup_{p\in P} \prol[E]{P}[p]$ is a vector bundle over $P$
and the vector bundle projection $\tau^E_P$ is just the projection
onto the first factor. The anchor of $\prol[E]{P}$ is the projection
onto the third factor, that is, the map
$\map{\rho^{\pi}}{\prol[E]{P}}{TP}$ given by $\rho^{\pi}(p,b,v)=v$.
The projection onto the second factor will be denoted by
$\map{\prol{\pi}}{\prol[E]{P}}{E}$, and it is a morphism of Lie
algebroids over $\pi$.  Explicitly, $\prol{\pi}(p,b,v)=b$.

An element $z\in\prol[E]{P}$ is said to be \emph{vertical} if it
projects to zero, that is $\prol{\pi}(z)=0$. Therefore it is of the
form $(p,0,v)$, with $v$ a $\pi$-vertical vector tangent to $P$ at
$p$.

Given local coordinates $(x^i,u^A)$ on $P$ and a local basis
$\{e_\alpha\}$ of sections of $E$, we can define a local basis
$\{\X_\alpha,\V_A\}$ of sections of $\prol[E]{P}$ by
\[
\X_\alpha(p)
=\Bigl(p,e_\alpha(\pi(p)),\rho^i_\alpha\pd{}{x^i}\at{p}\Bigr) \qquand
\V_A(p) = \Bigl(p,0,\pd{}{u^A}\at{p}\Bigr).
\]
If $z=(p,b,v)$ is an element of $\prol[E]{P}$, with $b=z^\alpha
e_\alpha$, then $v$ is of the form $v=\rho^i_\alpha
z^\alpha\pd{}{x^i}+v^A\pd{}{u^A}$, and we can write
\[
z=z^\alpha\X_\alpha(p)+v^A\V_A(p).
\]
Vertical elements are linear combinations of $\{\V_A\}$.

The anchor map $\rho^{\pi}$ applied to a section $Z$ of $\prol[E]{P}$
with local expression $Z = Z^\alpha\X_\alpha+V^A\V_A$ is the vector
field on $P$ whose coordinate expression is
\[
\rho^{\pi}(Z) = \rho^i_\alpha Z^\alpha \pd{}{x^i} + V^A\pd{}{u^A}.
\]

Next, we will see that it is possible to induce a Lie bracket
structure on the space of sections of $\prol[E]{P}$. For that, we say
that a section $\tilde{X}$ of $\tau^E_P: \prol[E]{P} \to P$ is
\emph{projectable} if there exists a section $X$ of $\tau: E \to M$
and a vector field $U\in \mathfrak{X}(P)$ which is $\pi$-projectable
to the vector field $\rho(X)$ and such that $\tilde{X}(p) =
(X(\pi(p)), U(p)),$ for all $p \in P$. For such a projectable section
$\tilde{X}$, we will use the following notation $\tilde{X} \equiv (X,
U)$. It is easy to prove that one may choose a local basis of
projectable sections of the space $\Sec{\tau^E_P}$.

The Lie bracket of two projectable sections $Z_1=(X_1, U_1)$ and
$Z_2=(X_2, U_2)$ is then given by
\[
\lcf Z_1,Z_2\rcf^{\pi}(p)=(p,\lcf X_1,X_2\rcf (x),[U_1,U_2](p)),
\qquad p \in P,\,\;\;\; x=\pi(p).
\]
Since any section of $\prol[E]{P}$ can be locally written as a linear
combination of projectable sections, the definition of the Lie bracket
for arbitrary sections of $\prol[E]{P}$ follows. In particular, the
Lie brackets of the elements of the basis are
\[
\lcf\X_\alpha,\X_\beta\rcf^{\pi}= C^\gamma_{\alpha\beta}\:\X_\gamma,
\qquad \lcf \X_\alpha,\V_B\rcf^{\pi}=0 \qquand \lcf
V_A,\V_B\rcf^{\pi}=0,
\]
and, therefore, the exterior differential is determined by
\begin{align*}
  &dx^i=\rho^i_\alpha \X^\alpha,
  &&du^A=\V^A,\\
  &d\X^\gamma=-\frac{1}{2}C^\gamma_{\alpha\beta}\X^\alpha\wedge\X^\beta,
  &&d\V^A=0,
\end{align*}
where $\{\X^\alpha,\V^A\}$ is the dual basis to $\{\X_\alpha,\V_A\}$.

The Lie algebroid $\prol[E]{P}$ is called the
\emph{prolongation of $E$ over }$\pi$ or the
\emph{${E}$-tangent bundle to $\pi$}.

\subsection{Lie groupoids}
In this section, we review the definition of a Lie groupoid and
present some basic facts generalities about them (see
\cite{Ma,Mac} for more details).  A \emph{groupoid} over a set $M$
is a set $G$ together with the following structural maps:
\begin{itemize}
\item A pair of maps $\alpha: G \to M$, the \emph{source}, and $\beta:
  G \to M$, the \emph{target}.  These maps define the set of
  composable pairs $$
  G_{2}=\set{(g,h) \in G \times
    G}{\beta(g)=\alpha(h)}.  $$
\item A \emph{multiplication} $m: G_{2} \to G$, to be denoted simply
  by $m(g,h)=gh$, such that
  \begin{itemize}
  \item $\alpha(gh)=\alpha(g)$ and $\beta(gh)=\beta(h)$,
  \item $g(hk)=(gh)k$.
  \end{itemize}
\item An \emph{identity section} $\epsilon: M \to G$ such that
  \begin{itemize}
  \item $\epsilon(\alpha(g))g=g$ and $g\epsilon(\beta(g))=g$.
  \end{itemize}
\item An \emph{inversion map} $i: G \to G$, to be denoted simply by
  $i(g)=g^{-1}$, such that
  \begin{itemize}
  \item $g^{-1}g=\epsilon(\beta(g))$ and $gg^{-1}=\epsilon(\alpha(g))$.
  \end{itemize}
\end{itemize}

A groupoid $G$ over a set $M$ will be denoted simply by the symbol $G
\rightrightarrows M$.

The groupoid $G \rightrightarrows M$ is said to be a \emph{Lie
  groupoid} if $G$ and $M$ are manifolds and all the structural maps
are differentiable with $\alpha$ and $\beta$ differentiable
submersions. If $G \rightrightarrows M$ is a Lie groupoid then $m$ is
a submersion, $\epsilon$ is an immersion and $i$ is a diffeomorphism.
Moreover, if $x \in M$, $\alpha^{-1}(x)$ (resp., $\beta^{-1}(x)$) will
be said the \emph{$\alpha$-fiber} (resp., the \emph{$\beta$-fiber}) of
$x$.

On the other hand, if $g \in G$ then the \emph{left-translation by $g
  \in G$} and the \emph{right-translation by $g$} are the
diffeomorphisms
$$
\begin{array}{lll}
  l_{g}: \alpha^{-1}(\beta(g)) \to
  \alpha^{-1}(\alpha(g))&; \; \;& h \to
  l_{g}(h) = gh, \\
  r_{g}: \beta^{-1}(\alpha(g)) \to
  \beta^{-1}(\beta(g))&; \; \;& h \to r_{g}(h) = hg.
\end{array}
$$
Note that $l_{g}^{-1} = l_{g^{-1}}$ and $r_{g}^{-1} = r_{g^{-1}}$.

A vector field $\tilde{X}$ on $G$ is said to be \emph{left-invariant}
(resp., \emph{right-invariant}) if it is tangent to the fibers of
$\alpha$ (resp., $\beta$) and $\tilde{X}(gh) =
(T_{h}l_{g})(\tilde{X}_{h})$ (resp., $\tilde{X}(gh) =
(T_{g}r_{h})(\tilde{X}(g))$, for $(g,h) \in G_{2}$.

Now, we will recall the definition of the \emph{Lie algebroid
associated with $G$}.

We consider the vector bundle $\tau: E_G \to M$, whose fiber at a
point $x \in M$ is $(E_G)_{x} = V_{\epsilon(x)}\alpha = Ker
(T_{\epsilon(x)}\alpha)$. It is easy to prove that there exists a
bijection between the space $\Sec{\tau}$ and the set of left-invariant
(resp., right-invariant) vector fields on $G$. If $X$ is a section of
$\tau: E_G \to M$, the corresponding left-invariant (resp.,
right-invariant) vector field on $G$ will be denoted $\lvec{X}$
(resp., $\rvec{X}$), where
\begin{equation}\label{linv}
\lvec{X}(g) = (T_{\epsilon(\beta(g))}l_{g})(X(\beta(g))),
\end{equation}
\begin{equation}\label{rinv}
\rvec{X}(g) = -(T_{\epsilon(\alpha(g))}r_{g})((T_{\epsilon
(\alpha(g))}i)( X(\alpha(g)))),
\end{equation}
for $g \in G$. Using the above facts, we may introduce a Lie
algebroid structure $(\lcf\cdot , \cdot\rcf, \rho)$ on $E_G$,
which is defined by
\begin{equation}\label{LA}
\lvec{\lcf X, Y\rcf} = [\lvec{X}, \lvec{Y}], \makebox[.3cm]{}
\rho(X)(x) = (T_{\epsilon(x)}\beta)(X(x)),
\end{equation}
for $X, Y \in \Sec{\tau}$ and $x \in M$ (for more details,
see \cite{CDW,Ma}).

Given two Lie groupoids $G \rightrightarrows M$ and $G'
\rightrightarrows M'$, a \emph{morphism of Lie groupoids} is a
smooth map $\Psi: G \to G'$ such that
\[
(g, h) \in G_{2} \Longrightarrow (\Psi(g), \Psi(h)) \in (G')_{2}
\]
and
\[
\Psi(gh) = \Psi(g)\Psi(h).
\]
A morphism of Lie groupoids $\Psi: G \to G'$ induces a smooth map
$\Phi_{0}: M \to M'$ in such a way that
\[
\alpha' \circ \Psi = \Phi_{0} \circ \alpha, \makebox[.3cm]{}
\beta' \circ \Psi = \Phi_{0} \circ \beta, \makebox[.3cm]{} \Psi
\circ \epsilon = \epsilon' \circ \Phi_{0},
\]
$\alpha$, $\beta$ and $\epsilon$ (resp., $\alpha'$, $\beta'$ and
$\epsilon'$) being the source, the target and the identity section
of $G$ (resp., $G'$).

Suppose that $(\Psi, \Phi_{0})$ is a morphism between the Lie
groupoids $G \rightrightarrows M$ and $G' \rightrightarrows M'$
and that $\tau: E_G \to M$ (resp., $\tau': E_{G'} \to M'$) is the
Lie algebroid of $G$ (resp., $G'$). Then, if $x \in M$ we may
consider the linear map $\Phi_x: (E_G)_{x}\to
(E_{G'})_{\Phi_{0}(x)}$ defined by
\begin{equation}\label{Amor}
\Phi_x(v_{\epsilon(x)}) = (T_{\epsilon(x)}\Psi)(v_{\epsilon(x)}),
\; \; \mbox{ for } v_{\epsilon(x)} \in A_{x}G.
\end{equation}
In fact, we have that the pair $(\Phi, \Phi_{0})$ is a morphism
between the Lie algebroids $\tau: E_G \to M$ and $\tau': E_{G'}
\to M'$ (see \cite{Ma,Mac}).

\paragraph{\textbf{Examples}}
We will present some examples of Lie groupoids.

1.- {\bf Lie groups}. Any Lie group $G$ is a Lie groupoid over
$\{\mathfrak{e} \}$, the identity element of $G$. The Lie algebroid
associated with $G$ is just the Lie algebra $\mathfrak{g}$ of $G$.

2.- {\bf The pair or banal groupoid}. Let $M$ be a manifold. The
product manifold $M \times M$ is a Lie groupoid over $M$ in the
following way: $\alpha$ is the projection onto the first factor
and $\beta$ is the projection onto the second factor; $\epsilon(x)
= (x, x)$, for all $x \in M$, $m((x, y), (y, z)) = (x, z)$, for
$(x, y), (y, z) \in M \times M$ and $i(x, y) = (y, x)$. $M \times
M \rightrightarrows M$ is called the \emph{pair or banal
groupoid}. If $x$ is a point of $M$, it follows that
\[
V_{\epsilon(x)}\alpha=\{0_x\}\times T_xM\subseteq T_xM\times T_xM
\cong T_{(x,x)}(M\times M).
\]
Thus, the linear maps
\[
\Phi_x:T_xM\to V_{\epsilon(x)}\alpha,\;\;\; v_x\to (0_x,v_x),
\]
induce an isomorphism (over the identity of $M$) between  the Lie
algebroids $\tau_M:TM\to M$ and $\tau:E_{M\times M}\to M.$

3.- {\bf The Lie groupoid associated with a fibration}. Let $\pi:
P \to M$ be a fibration, that is, $\pi$ is a surjective submersion
and denote by $G_{\pi}$ the subset of $P \times P$ given by
\[
G_{\pi} = \set{(p, p') \in P \times P}{\pi(p) = \pi(p')}.
\]
Then, $G_{\pi}$ is a Lie groupoid over $P$ and the structural maps
$\alpha_{\pi}$, $\beta_{\pi}$, $m_{\pi}$, $\epsilon_{\pi}$ and
$i_{\pi}$ are the restrictions to $G_{\pi}$ of the structural maps
of the pair groupoid $P \times P \rightrightarrows P$.

If $p$ is a point of $P$ it follows that
\[
V_{\epsilon_{\pi}(p)}\alpha_{\pi} = \set{(0_{p}, Y_{p}) \in T_{p}P
\times T_{p}P }{ (T_{p}\pi)(Y_{p}) = 0 }.
\]
Thus, if $(\tau_{P})_{|V\pi}: V\pi \to P$ is the vertical bundle to
$\pi$ then the linear maps
\[
(\Phi_{\pi})_{p}: V_{p}\pi \to V_{\epsilon_{\pi}(p)}
\alpha_{\pi}, \makebox[.3cm]{} Y_{p} \to (0_{p},
Y_{p})
\]
induce an isomorphism (over the identity of $M$) between the Lie
algebroids $(\tau_{P})_{|V\pi}: V\pi \to P$ and $\tau: E_{G_{\pi}}
\to P$.

4.- {\bf Atiyah or gauge groupoids}. Let $p: Q \to M$ be a
principal left $G$-bundle. Then, the free action $\Phi: G \times Q
\to Q$, $(g, q) \to \Phi(g, q) = gq$, of $G$ on $Q$ induces, in a
natural way, a free action $\Phi \times \Phi: G \times (Q \times
Q) \to Q \times Q$ of $G$ on $Q \times Q$ given by $(\Phi \times
\Phi)(g, (q, q')) = (gq, gq')$, for $g \in G$ and $(q, q') \in Q
\times Q$. Moreover, one may consider the quotient manifold $(Q
\times Q) / G$ which admits a Lie groupoid structure over $M$ with
structural maps given by
\[
\begin{array}{lcl}
\tilde{\alpha}: (Q \times Q) / G \to M &; \; \; &[(q,
q')] \to p(q),
\\
\tilde{\beta}: (Q \times Q) / G \to M &; \; \; &[(q,
q')] \to p(q'),
\\
\tilde{\epsilon}: M \to (Q \times Q) / G &; \; \;&
x\to [(q, q)], \; \; \mbox{ if } p(q) = x,
\\
\tilde{m}: ((Q \times Q) / G)_{2} \to (Q \times Q) / G
&; \; \;& ([(q, q')], [(gq', q'')]) \to [(gq, q'')],
\\
\tilde{i}: (Q \times Q) / G \to (Q \times Q) / G &; \;
\;& [(q, q')] \to [(q', q)].
\end{array}
\]
This Lie groupoid is called \emph{the Atiyah (gauge) groupoid
associated with the principal $G$-bundle $p: Q \to M$} (see
\cite{L}).

If $x$ is a point of $M$ such that $p(q) = x$, with $q \in Q$, and
$p_{Q \times Q}: Q \times Q \to (Q \times Q) / G$ is the canonical
projection then it is clear that
\[
V_{\tilde{\epsilon}(x)}\tilde{\alpha} = (T_{(q, q)}p_{Q \times
Q})(\{0_{q}\} \times T_{q}Q).
\]
Thus, if $\tau_{Q}|G: TQ / G \to M$ is the Atiyah algebroid
associated with the principal $G$-bundle $p: G \to M$ then the
linear maps
\[
(TQ / G)_{x} \to V_{\tilde{\epsilon}(x)}\tilde{\alpha} \; \; ; \;
\; [v_{q}] \to (T_{(q, q)}p_{Q \times Q})(0_{q}, v_{q}),\mbox{
with $v_{q} \in T_{q}Q$},
\]
induce an isomorphism (over the identity of $M$) between the Lie
algebroids $\tau: E_{(Q \times Q) / G} \to M$ and $\tau_{Q} | G:
TQ / G \to M$.

5.- {\bf Action Lie groupoids}. Let $G
\rightrightarrows M$ be a Lie groupoid and $f: M'
\to M$ be a smooth map. If $M' \mbox{$\;$}_f
\kern-3pt\times_\alpha G = \set{ (p, g) \in P
\times G}{ f(p) = \alpha (g)}$ then a \emph{right
action of $G$ on $f$} is a smooth map
\[
M' \mbox{$\;$}_f \kern-3pt\times_\alpha G \to M', \; \; (x', g)
\to x'g,
\]
which satisfies the following relations
$$\begin{array}{rcll}
f(x'g) &=& \beta(g), &\mbox{ for } (x', g) \in M' \mbox{$\;$}_f
\kern-3pt\times_\alpha G,\\
(x'g)h &=& x'(gh),    &\mbox{ for } (x', g) \in M' \mbox{$\;$}_f
\kern-3pt\times_\alpha G  \mbox{ and } (g, h) \in G_{2}, \mbox{ and }\\
x'\epsilon(f(x')) &=& x',   &\mbox{ for } x' \in M'.
\end{array}$$

Given such an action one constructs \emph{the action Lie groupoid}
$M' \mbox{$\;$}_f \kern-3pt\times_\alpha G$ over $M'$ by defining
\[
\begin{array}{lcl}
\tilde\alpha_{f}:  M' \mbox{$\;$}_f \kern-3pt\times_\alpha G
\to M' &; \; \; &(x', g) \to x',
\\
\tilde\beta_{f}: M' \mbox{$\;$}_f \kern-3pt\times_\alpha G
\to M' &; \; \;& (x', g) \to x'g,
\\
\tilde\epsilon_{f}: M' \to M' \mbox{$\;$}_f
\kern-3pt\times_\alpha G  &; \; \;& x' \to (x',
\epsilon(f(x'))),
\\
\tilde{m}_{f}: (M' \mbox{$\;$}_f \kern-3pt\times_\alpha G)_{2}
\to M' \mbox{$\;$}_f \kern-3pt\times_\alpha G &; \;
\;& ((x', g), (x'g, h)) \to (x', gh),
\\
\tilde{i}_{f}:  M' \mbox{$\;$}_f \kern-3pt\times_\alpha G
\to M' \mbox{$\;$}_f \kern-3pt\times_\alpha G &; \;
\;& (x', g) \to (x'g, g^{-1}).
\end{array}
\]
Now, if $x' \in M'$, we consider the map $x' \;\cdot :
\alpha^{-1}(f(x')) \to M'$ given by
\[
x' \cdot (g) = x'g.
\]
Then, if $\tau: E_G \to M$ is the Lie algebroid of $G$, the
$\R$-linear map $\map{\Phi}{\Sec{\tau}}{\mathfrak{X}(M')}$ defined
by
\[
\Phi(X)(x') = (T_{\epsilon(f(x'))}x' \;\cdot)(X(f(x'))), \; \;
\mbox{ for } X \in \Sec{\tau} \mbox{ and } x' \in M',
\]
induces an action of $E_G$ on $f: M' \to M$. In addition, the Lie
algebroid associated with the Lie groupoid $M' \mbox{$\;$}_f
\kern-3pt\times_\alpha G \rightrightarrows P$ is the action Lie
algebroid $E_G \ltimes f$ (for more details, see \cite{HM}).

6.- \textbf{The prolongation of a Lie groupoid over a fibration}.
Given a Lie groupoid $G\rightrightarrows M$ and a fibration $\pi: P \to
M$, we consider the set
\[
\Gprol{P} \equiv P \mbox{$\;$}_\pi
\kern-3pt\times_\alpha G \mbox{$\;$}_\beta
\kern-3pt\times_\pi P = \set{ (p, g, p') \in P
\times G \times P}{\pi(p) = \alpha(g), \; \;
\beta(g) = \pi(p') }.
\]
Then, $\Gprol{P}$ is a Lie groupoid over $P$
with structural maps given by
\[
\begin{array}{lcl}
\alpha^{\pi}: \Gprol{P} \to P &; \;
\; &(p, g, p') \to p,
\\
\beta^{\pi}: \Gprol{P}\to P &; \;
\;& (p, g, p') \to p',
\\
\epsilon^{\pi}: P \to \Gprol{P} &;
\; \;& p\to (p, \epsilon(\pi(p)), p),
\\
m^{\pi}: (\Gprol{P})_{2} \to
\Gprol{P} &; \; \;& ((p, g, p'), (p', h, p''))
\to (p, gh, p''),
\\
i^{\pi}: \Gprol{P} \to \Gprol{P}
&; \; \;& (p, g, p') \to (p', g^{-1},
p).
\end{array}
\]
$\Gprol{P}$ is called the \emph{prolongation of
G over $\pi: P \to M$}.

Now, denote by $\tau: E_G \to M$ the Lie algebroid of $G$, by
$E_{\Gprol{P}}$ the Lie algebroid of $\Gprol{P}$ and by
$\prol[E_G]{P}$ the prolongation of $\tau: E_G \to M$ over the
fibration $\pi$. If $p \in P$ and $m=\pi(p)$, then it follows that
\[
\left(E_{\Gprol{P}}\right)_p =  \set{ (0_{p}, v_{\epsilon(m)}, X_{p}) \in
T_{p}P \times (E_G)_{m} \times T_{p}P}{ (T_{p}\pi)(X_{p}) =
(T_{\epsilon(m)}\beta)(v_{\epsilon(m)})}
\]
and, thus, one may consider the linear
isomorphism
\begin{equation}\label{Isom}
(\Phi^{\pi})_{p}: (E_{\Gprol{P}})_p \to
\prol[E_G]{P}[p], \makebox[.3cm]{} (0_{p}, v_{\epsilon(m)}, X_{p})
\to (v_{\epsilon(m)}, X_{p}).
\end{equation}
In addition, one may prove that the maps $(\Phi^{\pi})_{p}$, $p
\in P$, induce an isomorphism $\Phi^{\pi}: E_{\Gprol{P}} \to
\prol[E_G]{P}$ between the Lie algebroids $E_{\Gprol{P}}$ and
$\prol[E_G]{P}$ (for more details, see \cite{HM}).

{\bf A particular case}. Next, suppose that $P = E_G$ and that the
map $\pi$ is just the vector bundle projection $\tau: E_G \to M$.
In this case,
\[
\Gprol{E_G} = E_G \mbox{$\;$}_\tau \kern-3pt\times_\alpha G
\mbox{$\;$}_\beta \kern-3pt\times_\tau E_G
\]
and we may define the map $\Theta: \Gprol{E_G} \to V\beta
\oplus_{G} V\alpha$ as follows
\[
\Theta(u_{\epsilon(\alpha(g))}, g,
v_{\epsilon(\beta(g))}) =
((T_{\epsilon(\alpha(g))}(r_{g} \circ
i))(u_{\epsilon(\alpha(g))}),
(T_{\epsilon(\beta(g))}l_{g})
(v_{\epsilon(\beta(g))})),
\]
for $(u_{\epsilon(\alpha(g))}, g, v_{\epsilon(\beta(g))}) \in
(E_G)_{\alpha(g)} \times G \times (E_G)_{\beta(g)}$. $\Theta$ is a
bijective map and
\[
\Theta^{-1}(X_{g}, Y_{g}) = ((T_{g}(i \circ
r_{g^{-1}}))(X_{g}), g,
(T_{g}l_{g^{-1}})(Y_{g})),
\]
for $(X_{g}, Y_{g}) \in V_{g}\beta \oplus V_{g}\alpha$. Thus, the
spaces $\Gprol{E_G}$ and $V\beta \oplus_{G} V\alpha$ may be
identified and, under this identification, the structural maps of
the Lie groupoid structure on $V\beta \oplus_{G} V\alpha$ are
given by
\[
\begin{array}{lcl}
\alpha^{\tau}: V\beta \oplus_{G} V\alpha \to E_G &; \;
\; &(X_{g}, Y_{g}) \to (T_{g}(i \circ
r_{g^{-1}}))(X_{g}),
\\
\beta^{\tau}: V\beta \oplus_{G} V\alpha \to E_G &; \;
\;& (X_{g}, Y_{g}) \to (T_{g}l_{g^{-1}})(Y_{g}),
\\
\epsilon^{\tau}: E_G \to V\beta \oplus_{G} V\alpha  &;
\; \;& v_{\epsilon(x)} \to
((T_{\epsilon(x)}i)(v_{\epsilon(x)}), v_{\epsilon(x)}),
\\
i^{\tau}: V\beta \oplus_{G} V\alpha
\to V\beta \oplus_{G} V\alpha &; \;
\;& (X_{g}, Y_{g}) \to
((T_{g}i)(Y_{g}), (T_{g}i)(X_{g})),
\end{array}
\]
and the multiplication $m^{\tau}: (V\beta
\oplus_{G} V\alpha)_{2} \to V\beta
\oplus_{G} V\alpha$ is
\[
m^{\tau}((X_{g}, Y_{g}), ((T_{g}(r_{gh}\circ
i))(Y_{g}), Z_{h})) = ((T_{g}r_{h})(X_{g}),
(T_{h}l_{g})(Z_{h})).
\]
This Lie groupoid structure was considered by Saunders \cite{Sau}.  We
remark that the Lie algebroid of $\Gprol{E_G} \cong V\beta \oplus_{G}
V\alpha \rightrightarrows E_G$ is isomorphic to the prolongation
$\prol[E_G]{E_G}$ of $E_G$ over $\tau: E_G \to M$.

\section{Mechanics on Lie algebroids}\label{Mechanics}

We recall that a symplectic section on a vector bundle
$\map{\pi}{F}{M}$ is a section $\omega$ of $\wedge^2\pi^*$ which is
regular at every point when it is considered as a bilinear form. By a
\emph{symplectic Lie algebroid} we mean a pair $(E,\omega)$ where
$\map{\tau}{E}{M}$ is a Lie algebroid and $\omega$ is a symplectic
section on the vector bundle $E$ satisfying the compatibility
condition $d\omega=0$, where $d$ is the exterior differential of $E$.

On a symplectic Lie algebroid $(E,\omega)$ we can define a dynamical
system for every function on the base, as in the standard case of a
tangent bundle. Given a function $H\in\cinfty{M}$ there is a unique
section $\sigma_H\in\sec{\tau}$ such that
\[
i_{\sigma_H}\omega=dH.
\]
The section $\sigma_H$ is said to be the \emph{Hamiltonian section}
defined by $H$ and the vector field $X_H=\rho(\sigma_H)$ is said to be
the \emph{Hamiltonian vector field} defined by $H$. In this way we get
the dynamical system $\dot{x}=X_H(x)$.

A symplectic structure $\omega$ on a Lie algebroid $E$ defines a
\emph{Poisson bracket} $\{\ ,\ \}^\omega$ on the base manifold $M$ as
follows. Given two functions $F,G\in\cinfty{M}$ we define the bracket
\[
\{F,G\}^\omega=\omega(\sigma_F,\sigma_G).
\]
It is easy to see that the closure condition $d\omega=0$ implies that
$\{\ ,\ \}^\omega$ is a Poisson structure on $M$. In other words, if
we denote by $\Lambda$ the inverse of $\omega$ as bilinear form, then
$\{F,G\}^\omega=\Lambda(dF,dG)$. The Hamiltonian dynamical system
associated to $H$ can be written in terms of the Poisson bracket as
$\dot{x}=\{x,H\}^\omega$.

By a \emph{symplectomorphism} between two symplectic Lie algebroids
$(E,\omega)$ and $(E',\omega')$ we mean an isomorphism of Lie
algebroids $(\Psi,\Psi_0)$ from $E$ to $E'$ such that
$(\Psi,\Psi_0)^*\omega'=\omega$.  In this case the base map $\Psi_0$
is a Poisson diffeomorphism, that is, it satisfies
$\Psi_0^*\{F',G'\}^{\omega'}=\{\Psi_0^*F',\Psi_0^*G'\}^{\omega}$, for
all $F',G'\in\cinfty{M'}$.

Sections~\ref{Lagrangian} and~\ref{Hamiltonian} describe two
particular and important cases of the above construction.

\subsection{Lagrangian Mechanics}\label{Lagrangian}
In~\cite{M} (see also \cite{LeMaMa}) a geometric formalism for
Lagrangian Mechanics on Lie algebroids was introduced. It is developed
in the prolongation $\TEE$ of a Lie algebroid $E$ over the vector
bundle projection $\tau: E\to M$. The canonical geometrical structures
defined on $\TEE$ are the following:
\begin{itemize}
\item The \emph{vertical lift} $\map{\xi\spV}{\tau^*E}{\TEE}$ given by
  $\xi\spV(a,b)=(a,0,b\spV_a)$, where $b\spV_a$ is the vector tangent
  to the curve $a+tb$ at $t=0$.
\item The \emph{vertical endomorphism} $\map{S}{\TEE}{\TEE}$ defined
  as follows:
  \[
  S(a,b,v)=\xi\spV(a,b)=(a,0,b_a\spV).
  \]
\item The \emph{Liouville section}, which is the vertical section
  corresponding to the Liouville dilation vector field:
  \[
  \Delta(a)=\xi\spV(a,a)=(a,0,a_a\spV).
  \]
\end{itemize}

We also mention that the \emph{complete lift} $X\spC$ of a section
$X\in\Sec{E}$ is the section of $\TEE$ characterized by the following
properties:
\begin{enumerate}
\item projects to $X$, i.e., $\prol{\tau}\circ X\spC=X\circ\tau$,
\item ${\mathcal L}_{X\spC}\hat{\mu}=\widehat{{\mathcal L}_X\mu}$,
\end{enumerate}
where by $\hat{\alpha}\in\cinfty{E}$ we denote the linear function
associated to $\alpha\in\Sec{E^*}$.

Given a Lagrangian function $L\in\cinfty{E}$ we define the
\emph{Cartan 1-section} $\theta_L\in\Sec{(\prol[E]{E})^*}$ and the
\emph{Cartan 2-section} $\omega_L\in\Sec{\wedge^2(\prol[E]{E})^*}$ and
the \emph{Lagrangian energy} $E_L\in C^{\infty}(E)$ as
\begin{equation}\label{Cartan-forms}
  \theta_L=S^*(dL) , \qquad
  \omega_L = -d\theta_L\qquand E_L=\mathcal{L}_{\Delta} L-L.
\end{equation}
If $(x^i, y^{\alpha})$ are local fibred coordinates on $E$,
$(\rho^i_{\alpha}, C^{\gamma}_{\alpha\beta})$ are the corresponding
local structure functions on $E$ and $\{ \X_{\alpha}, \V_{\alpha}\}$
is the corresponding local basis of sections of $\TEE$ then
\begin{equation}\label{endverlo}
  S\X_{\alpha} = \V_{\alpha}, \makebox[.3cm]{}
  S\V_{\alpha} = 0, \makebox[.3cm]{} \mbox{ for all
  } \alpha,
\end{equation}
\begin{equation}
  \label{Lioulo} \Delta = y^{\alpha}\V_{\alpha},
\end{equation}
\begin{equation}
  \label{omegaL} \omega_L
  =\pd{^2L}{y^\alpha\partial
    y^\beta}\X^\alpha\wedge \V^\beta
  +\frac{1}{2}\left(
    \pd{^2L}{x^i\partial y^\alpha}\rho^i_\beta-\pd{^2L}{x^i\partial
      y^\beta}\rho^i_\alpha+\pd{L}{y^\gamma}C^\gamma_{\alpha\beta}
  \right)\X^\alpha\wedge \X^\beta,
\end{equation}
\begin{equation}
  \label{EL} E_L=\pd{L}{y^\alpha}y^\alpha-L.
\end{equation}

From (\ref{endverlo}), (\ref{Lioulo}), (\ref{omegaL}) and (\ref{EL}),
it follows that
\begin{equation}
  \label{2.4'} i_{SX} \omega_{L} =
  -S^*(i_{X}\omega_{L}), \makebox[.3cm]{}
  i_{\Delta}\omega_{L} = -S^*(dE_{L}),
\end{equation}
for $X \in \Sec{{\prol[E]{E}}}$.

Now, a curve $t\to c(t)$ on $E$ is a solution of the
\emph{Euler-Lagrange equations} for $L$ if
\begin{itemize}
\item[-] $c$ is admissible (that is, $\rho(c(t))=\dot{m}(t)$, where
  $m=\tau\circ c$) and
\item[-] $\displaystyle{i_{(c(t), \dot{c}(t))}\omega_L(c(t))-dE_L
    (c(t))=0}$, for all $t$.
\end{itemize}
If $c(t)=(x^i(t), y^{\alpha}(t))$ then $c$ is a solution of the
Euler-Lagrange equations for $L$ if and only if
\begin{equation}
  \label{free-forces}\dot{x}^i=\rho_\alpha^iy^\alpha,\;\;\;\;
  \frac{d}{dt}\Bigl(\frac{\partial L}{\partial y^\alpha}\Bigr) +
  \frac{\partial L}{\partial y^\gamma}C_{\alpha\beta}^\gamma y^\beta
  -\rho_\alpha^i\frac{\partial L}{\partial x^i} =0.
\end{equation}

Note that if $E$ is the standard Lie algebroid $TM$ then the above
equations are the classical Euler-Lagrange equations for $L: TM \to
\R$.

On the other hand, the Lagrangian function $L$ is said to be
\emph{regular} if $\omega_L$ is a symplectic section, that is, if
$\omega_L$ is regular at every point as a bilinear form. In such a
case, there exists a unique solution $\Gamma_L$ verifying
\[
i_{\Gamma_L}\Omega_L-dE_L=0\; .
\]

In addition, using (\ref{2.4'}), it follows that
$i_{S\Gamma_{L}}\omega_{L} = i_{\Delta}\omega_{L} $ which implies
that $\Gamma_{L}$ is a \sode\ \emph{section}, that is,
$$S(\Gamma_{L})=\Delta,$$
or alternatively
$\prol{\tau}(\Gamma_L(a))=a$ for all $a\in E$.

Thus, the integral curves of $\Gamma_L$ (that is, the integral curves
of the vector field $\rho^{\tau}(\Gamma_L)$) are solutions of the
Euler-Lagrange equations for $L$. $\Gamma_L$ is called the
\emph{Euler-Lagrange section} associated with $L$.

From (\ref{omegaL}), we deduce that $L$ is regular if and only if the
matrix $\displaystyle{W_{\alpha\beta}=\frac{\partial^2 L}{\partial
    y^{\alpha}\partial y^{\beta}}}$ is regular. Moreover, the local
expression of $\Gamma_L$ is
\[
\Gamma_L=y^\alpha\X_\alpha+f^\alpha\V_\alpha ,
\]
where the functions $f^\alpha$ satisfy the linear equations
\begin{equation}
  \label{free-forces1} \pd{^2L}{y^\beta\partial
    y^\alpha}f^\beta+\pd{^2L}{x^i\partial y^\alpha}\rho^i_\beta
  y^\beta +\pd{L}{y^\gamma}C^\gamma_{\alpha\beta}y^\beta
  -\rho^i_\alpha\pd{L}{x^i} =0, \mbox{ for all } \alpha.
\end{equation}

\paragraph{\textbf{Examples}}

\

1.- {\bf Real Lie algebras of finite dimension}.  Let $\mathfrak{g}$
be a real Lie algebra of finite dimension and $L: \mathfrak{g}\to\R$
be a Lagrangian function. Then, the Euler-Lagrange equations for $L$
are just the well-known \emph{Euler-Poincar\'e equations} (see, for
instance, \cite{MaRa}).

2.- {\bf The tangent bundle}. Let $L: TM\to\R$ be a standard
Lagrangian function on the tangent bundle $TM$ of $M$. Then, the
resultant equations are the \emph{classical Euler-Lagrange equations}
for $L$.

3.- {\bf Foliations}. If the Lie algebroid is the tangent bundle of a
foliation ${\mathcal F}$ on $P$ then one recovers the classical
formalism of \emph{holonomic mechanics}.

4.- {\bf Atiyah algebroids}. Let $\tau_{Q}|G: TQ / G \to M$ be the
Atiyah algebroid associated with a principal $G$-bundle $p: Q\to M$
and $L:TQ|G\to \R$ be a Lagrangian function. Then, the Euler-Lagrange
equations for $L$ are just the \emph{Lagrange-Poincar\'e equations}
(see \cite{LeMaMa}).

5.- {\bf Action Lie algebroids}. Suppose that $\mathfrak{g}$ is a real
Lie algebra of finite dimension and that $\Phi: \mathfrak{g}\times
V^*\to V^*$ is a linear representation of $\mathfrak{g}$ on $V^*$. If
$L: \mathfrak{g}\times V^*\to \R$ is a Lagrangian function on the
action Lie algebroid $\mathfrak{g}\times V^*\to V^*$ then the
Euler-Lagrange equations for $L$ are just the so-called
\emph{Euler-Poincar\'e equations with advected parameters} or the
\emph{Euler-Poisson-Poincar\'e equations} (see~\cite{HMR}).

\subsection{Hamiltonian Mechanics}\label{Hamiltonian}
In this section, we discuss how the Hamiltonian formalism can be
developed for systems evolving on Lie algebroids (for more details,
see \cite{LeMaMa,Ma2}).

Let $\tau^*:E^*\to M$ be the vector bundle projection of the dual
bundle $E^*$ to $E$. Consider the prolongation $\prol[E]{E^*}$ of $E$
over $\tau^*,$
\begin{eqnarray*}
\prol[E]{E^*}&=&\set{(b,v)\in E\times TE^*}{ \rho(b)=(T\tau^*)(v)} \\
&=&\set{ (a^*, b, v)\in E^*\times E\times TE^* }{
\tau^*(a^*)=\tau(b), \rho(b)=(T\tau^*)(v)}.
\end{eqnarray*}
The canonical geometrical structures defined on $\prol[E]{E^*}$ are
the following:
\begin{itemize}
\item The \emph{Liouville section}
  $\Theta_E\in\Sec{(\prol[E]{E^*})^*}$ defined by
  \begin{equation}\label{Lio}
    \Theta_E(a^*)(b,v)=a^*(b).
  \end{equation}
\item The \emph{canonical symplectic section } $\Omega_E\in
  \Sec{\wedge^2(\prol[E]{E^*})^*}$ is defined by
  \begin{equation}\label{sym}
    \Omega_E=-d\Theta_E.
  \end{equation}
  where $d$ is the differential on the Lie algebroid $\prol[E]{E^*}$.
\end{itemize}
Take coordinates $(x^i, p_{\alpha})$ on $E^*$ and denote by
$\{{\mathcal Y}_{\alpha}, {\mathcal P}^{\beta}\}$ the local basis of
sections $\prol[E]{E^*}$, with
\[
{\mathcal Y}_{\alpha}(a^*)=\left( a^*, e_{\alpha}(\tau^*(a^*)),
  \rho^i_{\alpha}\frac{\partial}{\partial x^i}\right)\qquand {\mathcal
  P}^{\beta}(a^*)=\left( a^*, 0, \frac{\partial}{\partial
    p_{\alpha}}\right).
\]
In coordinates the Liouville and canonical symplectic sections are
written as
\[
\Theta_E=p_{\alpha}{\mathcal Y}^{\alpha}\qquand \Omega_E={\mathcal
  Y}^{\alpha}\wedge {\mathcal P}_{\alpha} +\frac{1}{2}p_{\gamma}
C^{\gamma}_{\alpha\beta}{\mathcal Y}^{\alpha}\wedge {\mathcal
  Y}^{\beta} ,
\]
where $\{{\mathcal Y}^{\alpha},{\mathcal P}_{\beta}\}$ is the dual
basis of $\{{\mathcal Y}_{\alpha},{\mathcal P}^{\beta}\}$.

Every function $H\in C^{\infty}(E^*)$ define a unique section
$\Gamma_H$ of $\prol[E]{E^*}$ by the equation
\[
i_{\Gamma_H}\Omega_E=dH ,
\]
and, therefore, a vector field $\rho^{\tau^*}(\Gamma_H)=X_H$ on $E^*$
which gives the dynamics. In coordinates,
\[
\Gamma_H=\frac{\partial H}{\partial p_{\alpha}}{\mathcal
  Y}_{\alpha}-\left( \rho^i_{\alpha} \frac{\partial H}{\partial x^i}+
  p_{\gamma} C^{\gamma}_{\alpha\beta} \frac{\partial H}{\partial
    p_{\beta}}\right) {\mathcal P}^{\alpha},
\]
and therefore,
\[
X_H=\rho^i_{\alpha}\frac{\partial H}{\partial
  p_{\alpha}}\frac{\partial}{\partial x^i}-\left(
  \rho^i_{\alpha} \frac{\partial H}{\partial x^i}+
  p_{\gamma} C^{\gamma}_{\alpha\beta}
  \frac{\partial H}{\partial p_{\beta}}\right)
\frac{\partial}{\partial p_{\alpha}}.
\]
Thus, the \emph{Hamilton equations} are
\begin{equation}\label{Hameq}
  \frac{d x^i}{d t}=\rho^i_{\alpha} \frac{\partial H}{\partial
    p_{\alpha}}\qquad \frac{d p_{\alpha}}{d t}=- \rho^i_{\alpha}
  \frac{\partial H}{\partial x^i}- p_{\gamma}
  C^{\gamma}_{\alpha\beta} \frac{\partial H}{\partial p_{\beta}}.
\end{equation}

The Poisson bracket $\{\ ,\ \}^{\Omega_E}$ defined by the canonical
symplectic section $\Omega_E$ on $E^*$ is the canonical Poisson
bracket, which is known to exists on the dual of a Lie
algebroid~\cite{CaWe}.

\paragraph{\textbf{Examples}}

\

1.- {\bf Real Lie algebras of finite dimension}.  If the Lie algebroid
$E$ is a real Lie algebra of finite dimension then the Hamilton
equations are just the well-known \emph{Lie-Poisson equations} (see,
for instance, \cite{MaRa}).

2.- {\bf The tangent bundle}. If $E$ is the standard Lie algebroid
$TM$ and $H: T^*M\to\R$ is a Hamiltonian function then the resultant
equations are the \emph{classical Hamilton equations} for $H$.

3.- {\bf Foliations}. If the Lie algebroid is the tangent bundle of a
foliation ${\mathcal F}$ then one recovers the classical formalism of
\emph{holonomic Hamiltonian mechanics}.

4.- {\bf Atiyah algebroids}. Let $\tau_{Q}|G: TQ / G \to M=Q|G$ be the
Atiyah algebroid associated with a principal $G$-bundle $p: Q\to M$
and $H:T^*Q|G\to \R$ be a Hamilton function. Then, the Hamilton
equations for $H$ are just the \emph{Hamilton-Poincar\'e equations}
(see \cite{LeMaMa}).

5.- {\bf Action Lie algebroids}. Suppose that $\mathfrak{g}$ is a real
Lie algebra of finite dimension, that $V$ is a real vector space of
finite dimension and that $\Phi: \mathfrak{g}\times V^*\to V^*$ is a
linear representation of $\mathfrak{g}$ on $V^*$. If $H:
\mathfrak{g}^*\times V^*\to \R$ is a Hamiltonian function on the
action Lie algebroid $\mathfrak{g}\times V^*\to V^*$ then the Hamilton
equations for $H$ are just the \emph{Lie-Poisson equations on the dual
  of the semidirect product of Lie algebras} $
\mathfrak{s}=\mathfrak{g}\circledS V$ (see \cite{HMR}).

\subsection[The Legendre transformation]{The Legendre transformation
  and the equivalence between the Lagrangian and Hamiltonian
  formalisms}\label{seccion3.4}

Let $L:E\to \R$ be a Lagrangian function and $\theta_L \in
\Sec{(\prol[E]{E})^*}$ be the Poincar\'{e}-Cartan $1$-section
associated with~$L$.

We introduce \emph{the Legendre transformation associated with $L$} as
the smooth map $Leg_L:E\to E^*$ defined by
\begin{equation}\label{LegL}
  Leg_L(a)(b)=\frac{d}{dt}L(a+tb)\big|_{t=0},
\end{equation}
for $a,b\in E_x,$ where $E_x$ is the fiber of $E$ over the point $x\in
M$. In other words $Leg_L(a)(b)=\theta_L(a)(z)$, where $z$ is a point
in the fiber of $\prol[E]{E}$ over the point $a$ such that ${\mathcal
  T}\tau(z)=b$.

The map $Leg_L$ is well-defined and its local expression in fibred
coordinates on $E$ and $E^*$ is
\begin{equation}\label{locLegL}
  Leg_L(x^i,y^\alpha)=(x^i,\frac{\partial
    L}{\partial y^\alpha}).
\end{equation}
From this local expression it is easy to prove that the Lagrangian
$L$ is regular if and only if $Leg_L$ is a local diffeomorphism.

The Legendre transformation induces a map $\prol{Leg_L}:\prol[E]{E}\to
\prol[E]{E^*}$ defined by
\begin{equation}\label{LLegL}
(\prol{Leg_L})(b,X_a)=(b,(T_aLeg_L)(X_a)),
\end{equation}
for $a,b\in E$ and $(a, b,X_a)\in \prol[E]{E}[a]\subseteq
E_{\tau(a)}\times E_{\tau(a)}\times T_aE,$ where $TLeg_L:TE\to TE^*$
is the tangent map of $Leg_L.$ Note that $\tau^*\circ Leg_L=\tau$ and
thus $\prol{Leg_L}$ is well-defined.

If we consider local coordinates on $\TEE$ (resp. $\prol[E]{E^*}$)
induced by the local basis $\{\X_{\alpha}, \V_{\alpha}\}$ (resp.,
$\{{\mathcal Y}_{\alpha}, {\mathcal P}^{\alpha}\}$) the  local
expression of $\prol{Leg_L}$ is
\begin{equation}\label{LLegloc}
\prol{Leg_L}(x^i,y^\alpha;
z^\alpha,v^\alpha)=(x^i,\frac{\partial L}{\partial y^\alpha};
z^\alpha,\rho_\beta^iz^\beta\frac{\partial^2 L}{\partial
x^i\partial y^\alpha} + v^\beta\frac{\partial^2 L}{\partial
y^\alpha\partial y^\beta}).
\end{equation}

The relationship between Lagrangian and Hamiltonian Mechanics is
given by the following result.

\begin{theorem}\cite{LeMaMa}\label{t3.2}
  The pair $(\prol{Leg_L},Leg_L)$ is a morphism between the Lie
  algebroids $(\prol[E]{E}, \lcf\cdot,\cdot\rcf^{\tau},\rho^\tau)$ and
  $(\prol[E]{E^*},\lcf\cdot,\cdot\rcf^{\tau^*},\rho^{\tau^*}).$
  Moreover, if $\theta_L$ and $\omega_L$ (respectively, $\Theta_E$ and
  $\Omega_E)$ are the Poincar\'{e}-Cartan $1$-section and $2$-section
  associated with $L$ (respectively, the Liouville $1$-section and the
  canonical symplectic section on $\prol[E]{E^*}$) then
  \begin{equation}\label{pullback}
    (\prol{Leg_L},Leg_L)^*(\Theta_E)=\theta_L,\;\;\;
    (\prol{Leg_L},Leg_L)^*(\Omega_E)=\omega_L.
  \end{equation}
\end{theorem}

In addition, in \cite{LeMaMa}, it is proved that if the Lagrangian $L$
is \emph{hyperregular}, that is, $Leg_L$ is a global diffeomorphism,
then $(\prol{Leg_L},Leg_L)$ is a symplectomorphism and the
Euler-Lagrange section $\Gamma_L$ associated with $L$ and the
Hamiltonian section $\Gamma_H$ are $(\prol{Leg_L},Leg_L)$-related,
that is,
\begin{equation}\label{Related}
  \Gamma_H\circ Leg_L=\prol{Leg_L}\circ \Gamma_L.
\end{equation}
Therefore, an admissible curve $a(t)$ on $\TEE$ is a solution of the
Euler-Lagrange equations if and only if the curve $\mu(t)=Leg_L(a(t))$
is a solution of the Hamilton equations.

\section{Nonholonomic Lagrangian systems on Lie
  algebroids}\label{nonlinear}

\subsection{Constrained Lagrangian systems}
In this section, we will discuss Lagrangian systems on a Lie algebroid
$\tau: E \to M$ subject to nonholonomic constraints.  The constraints
are real functions on the positions and generalized velocities which
constrain the motion to some submanifold $\cm$ of $E$. $\cm$ is the
\emph{constraint submanifold}.

We will assume that the constraints are purely nonholonomic, that is,
not all the generalized velocities are allowable, although all the
positions are permitted. So, we will suppose that
$\map{\pi=\tau|_\cm}{\cm}{M}$ is a fibration.

The constraints are linear if they are linear functions on $E$ or, in
more geometrical terms, if $\cm$ is a vector subbundle of $E$ over $M$
(Lagrangian systems subject to linear constraints were discussed in
\cite{CoMa,MeLa}).

In the general case, since $\pi$ is a fibration, the prolongation
$\prol[E]{\cm}$ is defined. We will denote by $r$ the dimension of the
fibers of $\map{\pi}{\cm}{M}$, that is
$r=\operatorname{dim}\cm-\operatorname{dim} M$.

Now, we define the bundle $\vd\to \cm$ of \emph{virtual displacements}
as the subbundle of $\tau^*E$ of rank $r$ whose fiber at a point
$a\in\cm$ is
\[
\vd[a]=\set{b\in E_{\tau(a)}}{b_a\spV\in T_a\cm}.
\]
In other words, the elements of $\vd$ are pairs of elements $(a,b)\in
E\oplus_{M} E$ such that
\[
\frac{d}{dt}\phi(a+tb)\at{t=0}=0,
\]
for every local constraint function $\phi$.

We also define the bundle of \emph{constraint forces} $\cf$ by
$\cf=S^*((\prol[E]{\cm})^\circ)$. Since $\pi$ is a fibration, the
transformation $S^*: (\prol[E]{\cm})^0 \to \cf$ defines an isomorphism
between the vector bundles $(\prol[E]{\cm})^0 \to \cm$ and $\cf \to
\cm$. Therefore, the rank of $\cf$ is $s = n - r$, where $n$ is the
rank of $E$.

Next, suppose that $L \in C^{\infty}(E)$ is a regular Lagrangian
function. Then, the pair $(L, \cm)$ is a \emph{constrained Lagrangian
  system}. Moreover, assuming the validity of a Chetaev's principle in
the spirit of that of standard Nonholonomic Mechanics (see
\cite{LeMa2}), the solutions of the system $(L, \cm)$ are curves $t
\to c(t)$ on $E$ such that:
\begin{itemize}
\item[--] $c$ is admissible (that is, $\rho(c(t))=\dot{m}(t)$, where
  $m=\tau\circ c$),
\item[--] $c$ is contained in $\cm$ and,
\item[--] $i_{(c(t), \dot{c}(t))}\omega_{L}(c(t)) - dE_{L}(c(t))
  \in \cf(c(t))$, for all $t$.
\end{itemize}

If $(x^i, y^{\alpha})$ are local fibred coordinates on $E$,
$(\rho_{\alpha}^i, C_{\alpha \beta}^{\gamma})$ are the
corresponding local structure functions of $E$ and
\[
\phi^A(x^i, y^{\alpha}) = 0, \; \; A = 1, \dots, s ,
\]
are the local equations defining $\cm$ as a submanifold of $E$, then
$\displaystyle\Bigl\{\frac{\partial \phi^A}{\partial y^{\alpha}}
\X^{\alpha}\Bigr\}_{A = 1, \dots, s}$ is a local basis of $\cf$.
Moreover, a curve $t \to c(t) = (x^i(t), y^{\alpha}(t))$ on $E$ is a
solution of the problem if and only if
\begin{equation}
  \label{LD-edo}
  \begin{aligned}
    &\dot{x}^i=\rho^i_{\alpha}y^{\alpha},\\
    &\frac{d}{dt}\left(\pd{L}{y^{\alpha}}\right)+\pd{L}{y^\gamma}C^\gamma_{\alpha
      \beta}y^{\beta} -\rho^i_{\alpha}\pd{L}{x^i} = \lambda_{A}
    \frac{\partial \phi^{A}}{\partial y^{\alpha}},\\
    &\phi^{A}(x^i, y^{\alpha})=0,
  \end{aligned}
\end{equation}
where $\lambda_A$ are the Lagrange multipliers to be determined.

These equations are called \emph{the Lagrange-d'Alembert equations for
  the constrained system $(L, \cm)$}. Note that if $E$ is the Lie
algebroid $TM$, then the above equations are just the standard
Lagrange-d'Alembert equations for the constrained system $(L, \cm)$.

Now, we will assume that the solution curves of the problem are
the integral curves of a section $\Gamma$ of $\prol[E]{E} \to E$.
Then, we may reformulate geometrically the problem as follows: we
look for a section $\Gamma$ of $\prol[E]{E} \to E$ such that
\begin{equation}\label{8.1}
\begin{array}{ll}
&(i_\Gamma\omega_L-dE_L)|_\cm\in\Sec{\cf}, \\[5pt]
&\Gamma|_\cm\in\Sec{\prol[E]{\cm}}.
\end{array}
\end{equation}
If $\Gamma$ is a solution of the above equations then, from
(\ref{2.4'}), we have that
\[
(i_{S\Gamma}\omega_{L} - i_{\Delta}\omega_{L})|_\cm = 0,
\]
which implies that $\Gamma$ is a \sode\ section along $\cm$, that
is, $(S\Gamma - \Delta)|\cm = 0$.

\subsection{Regularity, projection of the free dynamics and
  nonholonomic bracket}
We will discuss next the regularity of the constrained system $(L,
\cm)$ (the constrained system $(L, \cm)$ is \emph{regular} if
equations (\ref{8.1}) admit a unique solution $\Gamma$).

For this purpose, we will introduce two new vector bundles $F$ and
$\prol[\vd]{\cm}$ over $\cm$. The fibers of $F$ and $\prol[\vd]{\cm}$
at the point $a \in \cm$ are
$$\begin{array}{rcl}
  F_{a} & = & \omega_{L}^{-1}(\cf_{a}),\\[5pt]
  \prol[\vd]{\cm}[a] & = & \set{ z \in \prol[E]{\cm}[a] }{
    {\prol{\pi}(z)\in\vd[a]}}=\set{z\in \prol[E]{\cm}[a]}{S(z)\in
    \prol[E]{\cm}[a]}.
\end{array}$$
Then, one may prove the following result.
\begin{theorem}
\label{regular-nonl}\cite{CoLeMaMa} The following properties are
equivalent:
\begin{enumerate}
\item The constrained Lagrangian system $(L,\cm)$ is regular.
\item $\prol[E]{\cm} \cap F=\{0\}$.
\item $\prol[{\mathcal V}]{\cm} \cap(\prol[{\mathcal V}]{\cm})^\perp=\{0\}$.
\end{enumerate}
\end{theorem}
Here, the orthogonal complement is taken with respect to the
symplectic section~$\omega_{L}$.

Condition (ii) (or, equivalently, (iii)) in Theorem
\ref{regular-nonl} is locally equivalent to the regularity of the
matrix
\[
\Bigl({\mathcal C}^{AB} =  \frac{\partial \phi^A}{\partial
y^\alpha}W^{\alpha\beta}\frac{\partial \phi^B}{\partial
y^\beta}\Bigr)_{A,B=1,\dots ,s}
\]
where $(W^{\alpha \beta})$ is the inverse matrix of $\Bigl(W_{\alpha
\beta} = \displaystyle \frac{\partial^{2}L}{\partial
y^{\alpha}\partial y^{\beta}}\Bigr)$.

Thus, if $L$ is a Lagrangian function of mechanical type (that is,
$L(a)=\frac{1}{2}\Gc(a,a)-V(\tau(a))$, for all $a\in E$, with
$\Gc: E\times_M E\to \R$ a bundle metric on $E$  and $V: M\to \R$
a real function on $M$) then the constrained system $(L, \cm)$ is
always regular.

Now, assume that the constrained Lagrangian system $(L, \cm)$ is
regular. Then (ii) in Theorem \ref{regular-nonl} is equivalent to
$(\prol[E]{E})_{|\cm} = \prol[E]{\cm} \oplus F$ and we will denote
by $P$ and $Q$ the complementary projectors defined by this
decomposition
\[
P_{a}: \prol[E]{E}[a] \to \prol[E]{\cm}[a], \makebox[.3cm]{}
Q_{a}: \prol[E]{E}[a] \to F_a, \; \; \mbox{ for all } a \in \cm.
\]
Moreover, we have
\begin{theorem} \label{dym-nonl} \cite{CoLeMaMa} Let $(L, \cm)$ be a
  regular constrained Lagrangian system and let $\Gamma_{L}$ be the
  solution of the free dynamics, i.e., $i_{{\Gamma}_{L}}\omega_{L} =
  dE_{L}$. Then, the solution of the constrained dynamics is the
  \sode\ $\Gamma$ along $\cm$ obtained as follows
\[
\Gamma = P(\Gamma_{L}|\cm).
\]
\end{theorem}
On the other hand, $(3)$ in Theorem \ref{regular-nonl} is
equivalent to $(\prol[E]{E})_{|\cm}=\prol[\vd]{\cm}\oplus
(\prol[\vd]{\cm})^\perp$ and we will denote by $\bar{P}$ and
$\bar{Q}$ the corresponding projectors induced by this
decomposition, that is,
\[
\bar{P}_a:\prol[E]{E}[a]\to \prol[\vd]{\cm}[a],\;\;\;
\bar{Q}_a=\prol[E]{E}[a]\to (\prol[\vd]{\cm}[a])^\perp, \mbox{ for all
} a\in \cm.
\]
\begin{theorem}
\label{t8.3} \cite{CoLeMaMa} Let $(L,\cm)$ be a regular
constrained Lagrangian system, $\Gamma_L$ (respectively, $\Gamma$)
be the solution of the free (respectively, constrained) dynamics
and $\Delta$ be the Liouville section of $\prol[E]{E}\to E.$ Then,
$\Gamma=\bar{P}(\Gamma_L|{\cm})$ if and only if the restriction to
${\cm}$ of the vector field $\rho^{\tau}(\Delta)$ on $E$ is
tangent to $\cm$.
\end{theorem}
Note that if $\cm$ is a vector subbundle of $E$ then the vector field
$\rho^{\tau}(\Delta)$ is tangent to $\cm$. Therefore, using Theorem
\ref{t8.3}, it follows that
\begin{corollary}
  Under the same hypotheses as in Theorem \ref{t8.3} if $\cm$ is a
  vector subbundle of $E$ (that is, the constraints are linear) then
  $\Gamma=\bar{P}(\Gamma_L|{\cm})$.
\end{corollary}
Next, we will study the conservation of the Lagrangian energy for
the constrained Lagrangian system $(L, \cm)$.

Since $S^*: (\prol[E]{\cm})^0 \to \Psi$ is a vector bundle
isomorphism, it follows that there exists a unique section
$\alpha_{(L, \cm)}$ of $(\prol[E]{\cm})^0 \to \cm$ such that
\[
i_{Q(\Gamma_{L}|{\cm})}\omega_{L} = S^*(\alpha_{(L, \cm)}).
\]
Moreover, we have
\begin{theorem}[Conservation of the energy]
\label{conser-ener} \cite{CoLeMaMa}
If $(L, \cm)$ is a regular constrained Lagrangian system and
$\Gamma$ is the solution of the dynamics then ${\mathcal
L}_{\Gamma}(E_{L}|{\cm}) = 0$ if and only if $\alpha_{(L,
\cm)}(\Delta|{\cm}) = 0$. In particular, if the vector field
$\rho^{\tau}(\Delta)$ is tangent to $\cm$ then ${\mathcal
L}_{\Gamma}(E_{L}|{\cm}) = 0$.
\end{theorem}
Now, suppose that $f$ and $g$ are two smooth real functions on
$\cm$ and take arbitrary extensions to $E$ denoted by the same
letters. Then, we may define \emph{the nonholonomic bracket } of
$f$ and $g$ as follows
\[
\{f,g\}_{nh}=\omega_L(\bar{P}(X_f),\bar{P}(X_g))| {\cm},
\]
where $X_f$ and $X_g$ are the Hamiltonian sections on $\prol[E]{E}$
associated with $f$ and $g$, respectively.

The nonholonomic bracket is well-defined and, furthermore, it is not
difficult to prove the following result.
\begin{theorem}[The nonholonomic bracket]
  \cite{CoLeMaMa} The nonholonomic bracket is an almost-Poisson
  bracket, i.e., it is skew-symmetric and satisfies the Leibniz rule
  (it is a derivation in each argument with respect to the usual
  product of functions). Moreover, if $f \in C^{\infty}(\cm)$ is an
  observable, then the evolution $\dot{f}$ of $f$ is given by
  \[
  \dot{f} = \rho^{\tau}(R_{L})(f) + \{f, E_{L}| \cm \}_{nh},
  \]
  where $R_{L}$ is the section of $\prol[E]{\cm} \to \cm$ defined by
  $R_{L}= P(\Gamma_{L}| \cm) - \bar{P}(\Gamma_{L}| \cm)$. In particular,
  if the vector field $\rho^{\tau}(\Delta)$ is tangent to $\cm$ then
  \[
  \dot{f} = \{f, E_{L}| \cm \}_{nh}.
  \]
\end{theorem}

\subsection{Reduction}
Next, we will discuss a reduction process and its relation with Lie
algebroid epimorphisms. These results will be also valid for Lie
algebroids without nonholonomic constraints, simply taking ${\mathcal
  M}=E$ and ${\mathcal M}'=E'$ in the sequel.

Let $(L,\cm)$ be a regular constrained Lagrangian system on a Lie
algebroid $\tau:E\to M$ and $(L',\cm')$ be another constrained
Lagrangian system on a second Lie algebroid $\tau':E'\to M'.$
Suppose also that we have a fiberwise surjective morphism of Lie
algebroids $\Phi:E\to E'$ over a surjective submersion $\phi:M\to
M'$ such that:
\begin{enumerate}
\item[$i)$] $L=L'\circ \Phi,$
\item[$ii)$] $\Phi_{|\cm}:\cm\to \cm'$ is a surjective submersion and
\item[$iii)$] $\Phi(\vd[a])=\vd[\Phi(a)]',$ for all $a\in \cm.$
\end{enumerate}

Note that if $\cm$ and $\cm'$ are vector subbundles of $E$ and
$E'$, respectively, then conditions $i)$, $ii)$ and $iii)$ hold if
and only if
\[
L = L' \circ \Phi \; \; \mbox{ and } \; \; \Phi (\cm) = \cm'.
\]
In the general case, one may introduce the map $\prol[\Phi]{\Phi}:
\prol[E]{\cm} \to \prol[E']{\cm'}$ given by
\[
(\prol[\Phi]{\Phi})(b, v) = (\Phi(b), (T\Phi)(v)),
\makebox[.3cm]{} \mbox{ for } (b, v) \in \prol[E]{\cm},
\]
and we have that $\prol[\Phi]{\Phi}$ is a Lie algebroid epimorphism
over $\Phi$. In addition, the following results hold

\begin{theorem}[Reduction of the constrained dynamics]\cite{CoLeMaMa}
\label{t8.5} Let $(L, \cm)$ be a regular constrained Lagrangian
system on a  Lie algebroid $E$ and $(L',\cm')$ be a constrained
Lagrangian system on a second Lie algebroid $E'$. Assume that we
have a fiberwise surjective morphism of Lie algebroids $\Phi:E\to
E'$ over $\phi:M\to M'$ such that conditions $i)$, $ii)$ and
$iii)$ hold. Then:
\begin{enumerate}
\item
The constrained Lagrangian system $(L', \cm')$ is regular.
\item
If $\Gamma$ (respectively, $\Gamma'$) is the constrained dynamics
for $L$ (respectively, for $L'$) then $\prol[\Phi]{\Phi}\circ
\Gamma=\Gamma'\circ \Phi.$
\item
If $t \to c(t)$ is a solution of Lagrange-d'Alembert differential
equations for $L$ then $\Phi(c(t))$ is a solution of
Lagrange-d'Alembert differential equations for $L'.$
\end{enumerate}
\end{theorem}

\begin{theorem}[Reduction of the nonholonomic bracket]
\label{t8.6'}\cite{CoLeMaMa}
Under the same hypotheses as in Theorem \ref{t8.5}, we have that
\[
\{f'\circ \Phi,g'\circ
\Phi\}_{nh}=\{f',g'\}_{nh}'\circ \Phi\] for
$f',g'\in C^\infty(\cm'),$ where
$\{\cdot,\cdot\}_{nh}$ (respectively,
$\{\cdot,\cdot\}_{nh}'$) is the nonholonomic
bracket for the constrained system $(L,\cm)$
(respectively, $(L',\cm')$). In other words,
$\Phi:\cm\to \cm'$ is an almost-Poisson morphism
when on $\cm$ and $\cm'$ we consider the
almost-Poisson structures defined by the
corresponding nonholonomic brackets.
\end{theorem}

\paragraph{\textbf{Reduction by symmetries}}
Let $\phi:Q\to M$ be a principal $G$-bundle and $\tau:E\to Q$
be a Lie algebroid over $Q$. In addition, assume that we have an
action of $G$ on $E$ such that the quotient vector bundle $E/G$ is
defined and the set $\Sec{E}^G$ of equivariant sections of $E$ is a
Lie subalgebra of $\Sec{E}$. Then, $E'=E/G$ has a canonical Lie
algebroid structure over $M$ such that the canonical projection
$\Phi:E\to E'$ is a fiberwise bijective Lie algebroid morphism
over $\phi$ (see \cite{LeMaMa}).

Next, suppose that $(L,\cm)$ is a $G$-invariant regular
constrained Lagrangian system, that is, the Lagrangian function
$L$ and the constraint submanifold $\cm$ are $G$-invariant. Assume
also that $\cm$ is closed. Then, one may define a Lagrangian
function $L':E'\to \R$ on $E'$ such that
\[
L=L'\circ \Phi.
\]
Moreover, $G$ acts on $\cm$ and the set of orbits $\cm'=\cm/G$ of
this action is a quotient manifold, that is, $\cm'$ is a smooth
manifold and the canonical projection $\Phi_{|\cm}:\cm\to
\cm'={\cm }/{G}$ is a submersion. Thus, one may consider the
constrained Lagrangian system $(L',{\cm}')$ on $E'$.

Since the orbits of the action of $G$ on $E$ are the fibers of
$\Phi$ and $\cm$ is $G$-invariant, we deduce that
\[
V_a(\Phi)\subseteq T_a\cm, \mbox{ for all } a\in \cm,
\]
$V(\Phi)$ being the vertical bundle of $\Phi$. This implies that
$\Phi_{|\vd[a]}:{\vd[a]}\to \vd[\Phi(a)]'$ is a linear
isomorphism, for all $a\in \cm.$

Therefore, from Theorem \ref{t8.5}, we conclude that the
constrained Lagrangian system $(L',\cm')$ is regular and that
\[
\prol[\Phi]{\Phi}\circ \Gamma=\Gamma'\circ \Phi,
\]
where $\Gamma$ (resp., $\Gamma'$) is the constrained dynamics for $L$
(resp., $L'$). In addition, using Theorem \ref{t8.6'}, we obtain that
$\Phi: \cm \to \cm'$ is an almost-Poisson morphism when on $\cm$ and
$\cm'$ we consider the almost-Poisson structures induced by the
corresponding nonholonomic brackets.

\subsection{Example: a rolling ball on a rotating table}
We apply the results in this section to the case of a ball rolling
without sliding on a rotating table with constant angular velocity
\cite{BlKrMaMu,CLMM,CoLeMaMa,LM,NF}.  A (homogeneous) sphere of radius
$r>0$, unit mass $m=1$ and inertia $k^2$ about any axis, rolls without
sliding on a horizontal table which rotates with constant angular
velocity $\Omega$ about a vertical axis through one of its points.
Apart from the constant gravitational force, no other external forces
are assumed to act on the sphere.

Choose a Cartesian reference frame with origin at the center of
rotation of the table and $z$-axis along the rotation axis. Let
$(x,y)$ denote the position of the point of contact of the sphere with
the table. The configuration space for the sphere on the table is
$Q=\R^2\times SO(3)$, where $SO(3)$ may be parameterized by the
Eulerian angles $\theta,\varphi$ and $\psi$. The kinetic energy of the
sphere is then given by
\[
T=\frac{1}{2}(\dot{x}^2 + \dot{y}^2 + k^2(\dot\theta^2 + \dot\psi^2 +
2 \dot\varphi\dot\psi \cos \theta)) .
\]
With the potential energy being constant, we may put $V=0$. Thus, the
Lagrangian function $L$ is $T$ and the constraint equations are
\[
\begin{array}{lcr}
  \dot{x}-r\dot{\theta}\sin \psi + r \dot\varphi\sin \theta \cos
  \psi&=&-\Omega y,\\
  \dot{y} + r\dot\theta \cos\psi + r \dot{\varphi}\sin \theta \sin
  \psi&=&\Omega x.
\end{array}
\]
Since the Lagrangian function is of mechanical type, the constrained
system is regular. Note that the constraints are not linear and that
the restriction to the constraint submanifold $\cm$ of the Liouville
vector field on $TQ$ is not tangent to $\cm$. Indeed, the constraints
are linear if and only if $\Omega = 0$.

Next, following \cite{CLMM,CoLeMaMa}, we will consider local
coordinates
 $(\bar{x},\bar{y},\bar{\theta}, \bar{\varphi}, \bar{\psi};
 \pi_i)_{i=1,\dots ,5}$ on $TQ=T\R^2 \times T(SO(3))$, where
 \[
 \bar{x}=x,\;\;\; \bar{y}=y,\;\;\; \bar{\theta}=\theta,\;\;\;
 \bar{\varphi}=\varphi,\;\;\; \bar{\psi}=\psi,\]
 \[
 \pi_1=r\dot{x}+k^2\dot{q}_2,\;\;\;
 \pi_2=r\dot{y}-k^2\dot{q}_1,\;\;\; \pi_3=k^2\dot{q}_3,
 \]
 \[
 \pi_4=\frac{k^2}{(k^2+ r^2)}(\dot{x}-r\dot{q}_2 + \Omega y),\;\;\;
 \pi_5=\frac{k^2}{(k^2+ r^2)}(\dot{y}+r\dot{q}_1 - \Omega x),
\]
$(\dot{q}_1,\dot{q}_2,\dot{q}_3)$ are the quasi-coordinates
defined by
\[
\dot{q}_1=\omega_x,\;\;\; \dot{q}_2=\omega_y,\;\;\;
\dot{q}_3=\omega_z,
\]
and $\omega_{x}$, $\omega_{y}$ and $\omega_{z}$ are the components of
the angular velocity of the sphere.

Then, the constrained dynamics is the \sode\ $\Gamma$ along $\cm$
defined by
\begin{equation}\label{Gamma}
\begin{array}{rcl}
\Gamma = (P\Gamma_L|{\cm}) &=&(\dot{x}\displaystyle\frac{\partial
}{\partial \bar{x}} + \dot{y}\displaystyle\frac{\partial
}{\partial \bar{y}}+ \dot{\theta}\displaystyle\frac{\partial
}{\partial \bar{\theta}} +
\dot{\varphi}\displaystyle\frac{\partial }{\partial
\bar{\varphi}}+ \dot{\psi}\displaystyle\frac{\partial }{\partial
\bar{\psi}})| {\cm}
\\&=&(\dot{x}\displaystyle\frac{\partial }{\partial \bar{x}} +
\dot{y}\displaystyle\frac{\partial }{\partial \bar{y}}+
\dot{q_1}\displaystyle\frac{\partial }{\partial q_1} +
\dot{q_2}\displaystyle\frac{\partial }{\partial q_2}+
\dot{q_3}\displaystyle\frac{\partial }{\partial q_3})| {\cm}.
\end{array}
\end{equation}

On the other hand, when constructing the nonholonomic bracket on
$\cm$, we find that the only non-zero fundamental brackets are
\begin{equation}\label{Cornonh}
\begin{array}{ll}
  \{x,\pi_1\}_{nh}=r,&\kern-50pt\{y,\pi_2\}_{nh}=r,\\
  \{q_1,\pi_2\}_{nh}=-1,&\kern-50pt\{q_2,\pi_1\}_{nh}=1,\;\;\;\;\;\;\;\;
  \{q_3,\pi_3\}_{nh}=1,\\[5pt]
  \{\pi_1,\pi_2\}_{nh}=\pi_3,&\kern-50pt\{\pi_2,\pi_3\}_{nh}=\displaystyle\frac{k^2}{(k^2+r^2)}\pi_1
  + \displaystyle\frac{rk^2\Omega}{(k^2+
    r^2)}y,\\\{\pi_3,\pi_1\}_{nh}=\displaystyle\frac{k^2}{(k^2+r^2)}\pi_2
  - \displaystyle\frac{rk^2\Omega}{(k^2+ r^2)}x,
\end{array}
\end{equation}
in which the ``appropriate operational'' meaning has to be attached to
the quasi-coordinates $q_i$.

Thus, we have that
\[
\dot{f}=R_L(f) + \{f,L\}_{nh}, \mbox{ for $f\in C^\infty(\cm)$} ,
\]
where $R_L$ is the vector field on $\cm$ given by
\[
\begin{array}{lcl}
  R_L&=&\displaystyle(\frac{k^2\Omega}{(k^2+
    r^2)}(x\displaystyle\frac{\partial }{\partial
    y}-y\displaystyle\frac{\partial }{\partial x}) +
  \displaystyle\frac{r\Omega}{(k^2+
    r^2)}(x\displaystyle\frac{\partial }{\partial q_1} + y
  \frac{\partial
  }{\partial q_2} \\[5pt]&& + x(\pi_3-k^2\Omega)\displaystyle\frac{\partial }{\partial \pi_1}
  + y (\pi_3-k^2\Omega)\displaystyle\frac{\partial
  }{\partial \pi_2}-k^2(\pi_1x+\pi_2y)\displaystyle\frac{\partial
  }{\partial \pi_3}))| {\cm}.
\end{array}
\]
Note that $R_{L} = 0$ if and only if $\Omega = 0$.

Now, it is clear that $Q=\R^2\times SO(3)$ is the total space of a
trivial principal $SO(3)$-bundle over $\R^2$ and the bundle projection
$\phi:Q\to M=\R^2$ is just the canonical projection on the first
factor. Therefore, we may consider the corresponding Atiyah algebroid
$E'=TQ/SO(3)$ over $M=\R^2$.

One may prove that $E'$ is isomorphic to the real vector bundle $T\R^2
\times \R^3 \to \R^2$ in such a way that the anchor map $\rho': E'
\cong T\R^2 \times \R^3 \to T\R^2$ is just the canonical projection on
the first factor. Moreover, one may choose a global basis
$\{e'_i\}_{i=1, \ldots, 5}$ of $\Sec{E'}$ and the only non-zero
fundamental Lie brackets are
\[
\lcf e_4',e_3'\rcf'=e_5',\;\;\;\lcf e_5',e_4'\rcf'=e_3',\;\;\;
\lcf e_3',e_5'\rcf'=e_4'.
\]
We have that the Lagrangian function $L=T$ and the constraint
submanifold $\cm$ are $SO(3)$-invariant. Consequently, $L$ induces
a Lagrangian function $L'$ on $E'=TQ/SO(3)$ and the set of orbits
$\cm'=\cm/SO(3)$  is a submanifold of $E'=TQ/SO(3)$ in such a way
that the canonical projection $\Phi|\cm:\cm\to \cm'=\cm/SO(3)$ is
a surjective submersion.

Under the identification between $E'=TQ/SO(3)$ and $T\R^2\times \R^3,$
$L'$ is given by
\[
L'(x,y,\dot{x},\dot{y};\omega_1,\omega_2,\omega_3)=\frac{1}{2}(\dot{x}^2
+ \dot{y}^2) + \frac{k^2}{2} (\omega_1^2 + \omega_2^2 + \omega_3^2) ,
\]
where $(x,y,\dot{x},\dot{y})$ and $(\omega_1,\omega_2,\omega_3)$ are
the standard coordinates on $T\R^2$ and $\R^3$, respectively.
Moreover, the equations defining $\cm'$ as a submanifold of
$T\R^2\times \R^3$ are
\[
\dot{x}-r\omega_2 + \Omega y=0,\;\;\; \dot{y} + r\omega_1-\Omega
x=0.
\]
So, we have the constrained Lagrangian system $(L',\cm')$ on the
Atiyah algebroid $E'=TQ/SO(3)\cong T\R^2\times \R^3.$ Note that the
constraints are not linear and that if $\Delta'$ is the Liouville
section of the prolongation $\prol[E']{E'}$ then the restriction to
$\cm'$ of the vector field $(\rho')^{\tau'}(\Delta')$ is not tangent
to $\cm'$.

Now, if we put
\[\begin{array}{lll}
x'=x,&y'=y,&\\
\pi_1'=r\dot{x} +
k^2\omega_2,&\pi_2'=r\dot{y}-k^2\omega_1,&\pi_3'=k^2\omega_3,\\
\pi_4'=\displaystyle \frac{k^2}{(k^2 + r^2)}(\dot{x}-r\omega_2 +
\Omega y), &\pi_5'=\displaystyle \frac{k^2}{(k^2+r^2)}(\dot{y} + r
\omega_1 -\Omega x),
\end{array}
\]
then $(x',y',\pi_1',\pi_2',\pi_3',\pi_4',\pi_5')$ is a system of
global coordinates on $TQ/SO(3)\cong T\R^2\times \R^3.$ In these
coordinates the equations defining the submanifold $\cm'$ are
$\pi_4'=0$ and $\pi_5'=0$ and the canonical projection $\Phi:TQ\to
TQ/SO(3)$ is given by
\begin{equation}\label{Phi}
\Phi(\bar{x},\bar{y},\bar{\theta},\bar{\varphi},\bar{\psi};
\pi_1,\pi_2,\pi_3,\pi_4,\pi_5)=(\bar{x},\bar{y};{\pi}_1, {\pi}_2,
{\pi}_3, {\pi}_4, {\pi}_5).
\end{equation}

Thus, if $\Gamma'$ is the constrained dynamics for the system $(L',
\cm')$, it follows that (see (\ref{Gamma}))
\[
(\rho')^{\tau'}(\Gamma')=(\dot{x}'\frac{\partial }{\partial x'} +
\dot{y}'\frac{\partial }{\partial y'})| {\cm'}.
\]

On the other hand, from (\ref{Cornonh}),
(\ref{Phi}) and Theorem \ref{t8.6'}, we deduce
that the only non-zero fundamental nonholonomic
brackets for the system $(L',\cm')$ are
\[
\begin{array}{lll}
  \{x',\pi_1'\}'_{nh}=r,& \{y',\pi_2'\}'_{nh}=r,&\\
  \{\pi_1',\pi_2'\}_{nh}'=\pi_3',&\{\pi_2',\pi_3'\}_{nh}' =
  \displaystyle\frac{k^2}{(k^2+
    r^2)}\pi_1' + \displaystyle\frac{rk^2\Omega}{(k^2+ r^2)}y',&\\
  \{\pi_3',\pi_1'\}_{nh}'=\displaystyle\frac{k^2}{(k^2 + r^2)}
  \pi_2'- \displaystyle\frac{rk^2\Omega}{(k^2+ r^2)}x'.&&
\end{array}
\]
Therefore, we have that
\[
\dot{f}'=(\rho')^{\tau'}(R_{L'})(f') + \{f',L'\}_{nh}',\mbox{ for
} f'\in C^\infty(\cm'),
\]
where $(\rho')^{\tau'}(R_{L'})$ is the vector field on $\cm'$ given by
\[
\begin{array}{rcl}
  (\rho')^{\tau'}(R_{L'})&=&\{ \displaystyle\frac{k^2\Omega}{k^2+
    r^2}(x'\displaystyle\frac{\partial }{\partial
    y'}-y'\displaystyle\frac{\partial }{\partial x'}) +
  \displaystyle\frac{r\Omega}{(k^2 +
    r^2)}(x'(\pi_3'-k^2\Omega)\displaystyle\frac{\partial }{\partial
    \pi_1'} \\[5pt]
  &&+ y' (\pi_3'-k^2\Omega)\displaystyle\frac{\partial }{\partial
    \pi_2'}-k^2(\pi_1' x' + \pi_2' y')\displaystyle\frac{\partial
  }{\partial \pi_3'})\}_{|\cm'}.
\end{array}
\]

\subsection{Hamiltonian formalism}
Let $(L, \cm)$ be a constrained Lagrangian system on a Lie algebroid
$E$ and assume that the Lagrangian function L is hyperregular. Then,
since the $Leg_L$ is a diffeomorphism then, it is clear that one may
develop a Hamiltonian formalism which is equivalent, via the Legendre
transformation, to the Lagrangian formalism.

\section{Mechanical control systems on Lie algebroids}\label{control}

\subsection{General control systems on Lie algebroids}
\label{se:general-control-systems}

Consider a Lie algebroid $\map{\tau}{E}{M}$, with anchor map
$\map{\rho}{E}{TM}$. Let $\{\sigma, \eta_1,\dots,\eta_k \}$ be
sections of $E$. A \emph{control problem on the Lie algebroid
  $\map{\tau}{E}{M}$ with drift section $\sigma$ and input sections
  $\eta_1,\dots,\eta_k$} is defined by the following equation on $M$,
\begin{equation}\label{eq:control-system}
  \dot{m} (t) = \rho\Bigl(\sigma (m(t)) + \sum_{i=1}^k u_i(t)\eta_i
  (m(t))\Bigr)  ,
\end{equation}
where $u=(u_1,\dots, u_k) \in U$, and $U$ is an open set of $\real^k$
containing $0$. The function $ t \mapsto u(t) = (u_1(t), \ldots,
u_k(t))$ belongs to a certain class of functions of time, denoted by
$\mathcal{U}$, called the \emph{set of admissible controls}.  For our
purposes, we may restrict the admissible controls to be the piecewise
constant functions with values in $U$. Notice that the trajectories of
the control system are admissible curves of the Lie algebroid, and
therefore they must lie on a leaf of $E$.  It follows that if $E$ is
not transitive, then there are points that cannot be connected by
solutions of any control system defined on such a Lie algebroid. In
particular, the system~\eqref{eq:control-system} cannot be locally
accessible at points $m \in M$ where $\rho$ is not surjective.  Since
the emphasis here is put on the controllability analysis, without loss
of generality we will restrict our attention to locally transitive Lie
algebroids.

Denoting by $f=\rho(\sigma)$ and $g_i=\rho(\eta_i)$, $i \in
{1,\dots,k}$, we can rewrite the system~\eqref{eq:control-system}
as
\begin{equation}\label{eq:standard}
  \dot{m} (t) = f (m(t)) + \sum_{i=1}^k u_i(t) g_i (m(t))  ,
\end{equation}
which is a standard nonlinear control system on $M$ affine in the
inputs~\cite{HN-AJvdS:90}.  Here we make use of the additional
geometric structure provided by the Lie algebroid in order to carry
over the analysis of the controllability properties of the control
system~\eqref{eq:control-system}. We refer to~\cite{HN-AJvdS:90} for a
comprehensive discussion of the notions of reachable sets,
accessibility algebra and computable accessibility tests.

\begin{definition}
  The \emph{accessibility algebra~$\D$ of the control
    system~\eqref{eq:control-system} in the Lie algebroid} is the
  smallest subalgebra of $\Sec{E}$ that contains the sections
  $\{\sigma,\eta_1,\dots,\eta_k\}$.
\end{definition}

Using the Jacobi identity, one can deduce that any element of
accessibility algebra~$\D$ is a linear combination of repeated Lie
brackets of sections of the form
\[
\lcf\zeta_l,\lcf\zeta_{l-1},\lcf\dots,\lcf\zeta_2,\zeta_1\rcf\dots\rcf\rcf\rcf
,
\]
where $\zeta_i \in \{\sigma,\eta_1,\dots,\eta_k\}$, $1 \le i \le
l$ and $l \in \natural$.

\begin{definition}
  The \emph{accessibility subbundle in the Lie algebroid}, denoted by
  $\invclos{\{\sigma,\eta_1,\dots,\eta_k\}}$, is the vector subbundle
  of $E$ generated by the accessibility algebra~$\D$,
  \[
  \invclos{\{\sigma,\eta_1,\dots,\eta_k\}} = \spn \left\{ \zeta(m) \;
    | \; \zeta \; \text{section of $E$ in} \; \D \right\} , \quad m
  \in M .
  \]
\end{definition}
If the dimension of $\invclos{\{\sigma,\eta_1,\dots,\eta_k\}}$ is
constant, then $\invclos{\{\sigma,\eta_1,\dots,\eta_k\}}$ is the
smallest Lie subalgebroid of $E$ that has
$\{\sigma,\eta_1,\ldots,\eta_k\}$ as sections.

\subsection{Mechanical control systems}

Let $\map{\tau}{E}{M}$ be a Lie algebroid, let $\nabla$ be a
connection on $E$, and let $\{\eta,\eta_1,\dots,\eta_k\}$ be
sections of $E$.  A \emph{mechanical control system on the Lie
algebroid $\map{\tau}{E}{M}$} is defined by the following equation
\begin{align}\label{eq:connection-control}
  \nabla_{a(t)}a(t) + \eta (m(t)) = \sum_{i=1}^k u_i (t) \eta_i (m(t))
  .
\end{align}
We will often refer to $\eta$ as the potential energy term in
equations~\eqref{eq:connection-control}. Associated with this
equation, there is always a control system on the Lie algebroid
$\prol[E]{E} \maparrow E$ given by
\begin{equation}\label{eq:CS}
  \dot{a}(t) = \rho^{\tau} \Bigl ( (\Gamma_{\nabla} - \eta\spV )(a(t)) +
  \sum_{i=1}^k u_i(t) \eta_i\spV (a(t))\Bigr),
\end{equation}
where $\eta\spV$ (resp. $\eta_i\spV$) denotes the vertical lift of
$\eta$ (resp. $\eta_i$) and $\Gamma_{\nabla}$ is the \sode\
section associated with $\nabla$. $\Gamma_{\nabla}$ is locally
given by (see \cite{CoMa})
\[
\Gamma_{\nabla}=y^{\alpha} \X_{\alpha}-\frac{1}{2}\left(
\Gamma^{\alpha}_{\beta\gamma}+\Gamma^{\alpha}_{\gamma\beta}\right)
y^{\beta}y^{\gamma}  \V_{\alpha}.
\]

There are two distinguished families within the class of mechanical
control systems. We introduce them next.

\paragraph{\bf Mechanical control systems} Consider a Lagrangian system
with $\map{L}{E}{\real}$ of the form
\[
L(a)=\frac{1}{2} \Gc(a,a) - V \circ \tau (a) , \quad a\in E,
\]
with $\map{\Gc}{E\times_M E}{\real}$ a bundle metric on $E$ and $V$ a
function on $M$. This Lagrangian function gives rise to the
Euler-Lagrange equations as explained in Section~\ref{Lagrangian}.

Consider now the situation when the Lagrangian system is subject
to some external forces, represented by a collection
$\{\theta_1,\dots,\theta_k\}$ of sections of $E^*$.  Denote by
$\{\eta_1,\dots,\eta_k\}$ the input sections of $E$ determined by
the control forces $\{\theta_1,\dots,\theta_k\}$ via the metric,
i.e., $\theta_i(X)=\Gc(\eta_i, X)$ for all $X\in \Sec{E}$.
If $\Gamma_{\nabla^{\Gc}}$ denotes the \sode\ section associated
with the Levi-Civita connection $\nabla^\Gc$, the controlled
Euler-Lagrange equations can be written as
\begin{equation}\label{eq:LCS}
  \dot{a}(t) = \rho^{\tau} \Bigl (\Gamma_{\nabla^\Gc} (a(t)) - (\grad_\Gc
  V) \spV (a(t))+ \sum_{i=1}^k u_i(t) \eta_i\spV (a(t))\Bigr).
\end{equation}
Here $\grad_\Gc V$ is the section of $E$ characterized by
$\Gc(\grad_\Gc V, X)=dV(X)$ for $X\in \Sec{E}$. Note that system
(\ref{eq:LCS}) is a control problem on the Lie algebroid $\prol[E]{E}
\maparrow E$ as defined in Section~\ref{se:general-control-systems}.
Locally, the equations can be written as
\begin{align*}
  \dot{x}^i & =\rho^i_\alpha y^\alpha , \\[-7pt]
  \dot{y}^\alpha &= -\frac{1}{2} \left( \CC^\alpha_{\beta \gamma} (x)
    +\CC^\alpha_{\gamma\beta} (x) \right) y^\beta y^\gamma -
  \Gc^{\alpha \beta} \rho^i_{\beta} \pd{V}{x^i} +\sum_{i=1}^k
  u_i(t)\eta^\alpha_i(x),
\end{align*}
where $(\Gc_{\alpha\beta})$ are the components of
the metric $\Gc$ and $(\Gc^{\alpha\beta})$ is the
inverse matrix of $(\Gc_{\alpha\beta})$.

 Alternatively, one can describe the dynamical behavior of
the mechanical control system by means of an equation on $E$ via
the covariant derivative.  An admissible curve $a: t \mapsto a(t)$
is a solution of the system~\eqref{eq:LCS} if and only if
\begin{align}\label{eq:eqs-motion-connection}
  \nabla^{\Gc}_{a(t)}a(t) + \grad_\Gc V(m(t)) = \sum_{i=1}^k u_i(t)
  \eta_i(m(t))  .
\end{align}
This equation corresponds to a mechanical control
system~\eqref{eq:connection-control} with connection $\nabla =
\nabla^{\Gc}$ and sections $\{\grad_\Gc V, \eta_1, \dots, \eta_k\}$.

\paragraph{\bf Mechanical control systems with constraints}
Assume a mechanical control system with data
$(\Gc,V,\{\theta_1,\dots,\theta_k\})$ is subject to the constraints
determined by a subbundle $D$ of $E$. Consider the orthogonal
decomposition $E=D\oplus D^{\perp}$ an the associated orthogonal
projectors $\map{P}{E}{D}$, $\map{Q}{E}{D^\perp}$. Then, one can write
the controlled Lagrange-d'Alembert equations as
\[
P(\nabla^\Gc_{a(t)}a(t)) + P(\grad_\Gc V(m(t))) = \sum_{i=1}^k u_i
(t) P(\eta_i(m(t))) , \qquad Q(a)=0 .
\]
In terms of the constrained connection $\cnabla_\sigma\eta =
P(\nabla^\Gc_\sigma\eta)+\nabla^\Gc_\sigma(Q\eta)$, with $\sigma,\eta
\in \Sec{E}$, the controlled equations can be rewritten as
$\cnabla_{a(t)}a(t) + P(\grad_\Gc V(m(t))) = \sum_{i=1}^k u_i (t)
P(\eta_i(m(t)))$, $Q(a)=0$.  Since the forcing terms coming from the
potential and the inputs belong to $D$, the solutions of the total
controlled dynamics initially belonging to $D$ also remain in $D$.  As
a consequence, an admissible curve $a: t \mapsto a(t)$ is a solution
of the system~\eqref{eq:LACS} if and only if
\begin{align}\label{eq:eqs-motion-connection-nh}
  \cnabla_{a(t)}a(t) + P(\grad_\Gc V(m(t))) = \sum_{i=1}^k u_i (t)
  P(\eta_i(m(t)))  ,\qquad a_0 \in D  .
\end{align}
This equation corresponds to a mechanical control
system~\eqref{eq:connection-control} with connection $\nabla =
\cnabla$ and sections $\{P(\grad_\Gc V), P(\eta_1), \dots,
P(\eta_k)\}$.

Note that one can write the controlled dynamics as a control
system on the Lie algebroid $\prol[E]{E} \maparrow E$,
\begin{equation}\label{eq:LACS}
  \dot{a}(t) = \rho^{\tau} \Bigl (\Gamma_{\cnabla} (a(t)) - P(\grad_\Gc
  V) \spV (a(t))+ \sum_{i=1}^k u_i(t) P(\eta_i)\spV (a(t))\Bigr).
\end{equation}
The coordinate expression of these equations is greatly simplified
if we take a basis $\{e_\alpha\}=\{e_a,e_A\}$ of $E$ adapted to
the orthogonal decomposition $E=D\oplus D^\perp$, i.e., $D = \spn
\{e_a\}$, $\D^\perp = \spn \{e_A\}$. Denoting by
$(y^\alpha)=(y^a,y^A)$ the induced coordinates, the constraint
equations $Q(a)=0$ just read $y^A=0$. The controlled
equations~\eqref{eq:eqs-motion-connection-nh} are then
\begin{align*}
  \dot{x}^i & = \rho^i_a y^a , \\
  \dot{y}^a & = - \frac{1}{2}\SC^a_{bc}y^by^c - \Gc^{a\beta}
  \rho^i_\beta\pd{V}{x^i} + \sum_{i=1}^k u_i (t) P(\eta_i)^a , \\
  y^A &= 0.
 \end{align*}
where $\SC^a_{bc}=\Gamma^b_{ca}+\Gamma^c_{ba}$
are the components of the symmetric product.

\subsection{Accessibility and controllability
  notions}\label{se:notions}

Here we introduce the notions of accessibility and controllability
that are specialized to mechanical control systems on Lie
algebroids. Let $m \in M$ and consider a neighborhood $V$ of $m$
in $M$. Define the set of reachable points in the base manifold
$M$ starting from $m$ as
\begin{multline*}
  \R_M^V (m, T) = \left\{ m' \in M \, \right. | \left. \exists u \in
    \mathcal{U} \; \hbox{defined on $[0,T]$ such that the evolution
      of~\eqref{eq:CS}} \right. \\
  \left. \hbox{for $a(0)=0_{m}$ satisfies} \; \tau(a(t)) \in V , \, t
    \in [0,T] \, \hbox{and} \; \tau(a(T))=m' \right\} .
\end{multline*}
Alternatively, one may write $\R_M^V (m, T) = \tau
(\R_E^{\tau^{-1}(V)} (0_m, T))$. Denote
\[
\R_M^V (m,\le T) = \bigcup_{t \le T} \R_M^V (m, t) .
\]

\begin{definition}\label{dfn:base-controllable}
  The system~\eqref{eq:CS} is \emph{locally base accessible from $m$}
  (respectively, \emph{locally base controllable from $m$}) if
  $\R_M^V(m,\le T)$ contains a non-empty open set of $M$
  (respectively, $\R_M^V(m,\le T)$ contains a non-empty open set of
  $M$ to which $m$ belongs) for all neighborhoods $V$ of $m$ and all
  $T>0$. If this holds for any $m \in M$, then the system is called
  \emph{locally base accessible} (respectively, \emph{locally base
    controllable}).
\end{definition}

In addition to the notions of base accessibility and base
controllability, we shall also consider full-state accessibility
and controllability starting from points of the form $0_m \in E$,
$m \in M$ (note that full-state is meant here with regards to $E$,
not to $TM$).

\begin{definition}\label{dfn:zero-controllable}
  The system~\eqref{eq:CS} is \emph{locally accessible from $m$ at
    zero} (respectively, \emph{locally controllable from $m$ at zero})
  if $\R_E^W(0_{m},\le T)$ contains a non-empty open set of $E$
  (respectively, $\R_E^W(0_{m},\le T)$ contains a non-empty open set
  of $E$ to which $0_{m}$ belongs) for all neighborhoods $W$ of
  $0_{m}$ in $E$ and all $T>0$. If this holds for any $m \in M$, then
  the system is called \emph{locally accessible at zero}
  (respectively, \emph{locally controllable at zero}).
\end{definition}

The relevance of the above definitions stems from the fact that,
frequently, one needs to control a system by starting at rest.
Nevertheless it is important to notice that not every equilibrium
point at $m$ corresponds to the point $0_m$.  Finally, we also
introduce the notion of accessibility and controllability with
regards to a manifold.

\begin{definition}\label{dfn:fiber-controllable}
  Let $\map{\psi}{M}{N}$ be an open mapping.  The
  system~\eqref{eq:CS} is \emph{locally base accessible from $m$ with
    regards to $N$} (respectively, \emph{locally base controllable
    from $m$} with regards to $N$) if $\psi(\R_M^V(m,\le T))$ contains
  a non-empty open set of $N$ (respectively, $\psi(\R_M^V(m,\le T))$
  contains a non-empty open set of $N$ to which $\psi(m)$ belongs) for
  all neighborhoods $V$ of $m$ and all $T>0$.  If this holds for any
  $m \in M$, then the system is called \emph{locally base accessible
    with regards to $N$} (respectively, \emph{locally base
    controllable with regards to $N$}).
\end{definition}

Note that base accessibility and controllability with regards to
$M$ with $\Id{M}:{M}\to {M}$ corresponds to the notions of base
accessibility and controllability (cf.
Definition~\ref{dfn:base-controllable}). Moreover, if the system
is base accessible, then it is base accessible with regards to
$N$. The analogous implication for base controllability also holds
true.

\subsection{The structure of the control Lie
  algebra}\label{se:structure}

The aim of this section is to show that the analysis of the
structure of the control Lie algebra of affine connection control
systems carried out in~\cite{ADL-RMM:95c} can be further extended
to control systems defined on a Lie algebroid. The enabling
technical notion exploited here is that of homogeneity.

Let $B$ be a Lie bracket formed with sections of the family $\famX
= \{ \Gamma_\nabla,\eta_1\spV,\dots,\eta_k\spV,\eta\spV \}$.  The
\emph{degree} of $B$ is the number of occurrences of all its
factors, and is therefore given by $\delta (B) = \delta_0(B) +
\delta_1 (B) + \dots + \delta_k(B)$, where $\delta_0(B)$,
$\delta_i(B)$, $i \in \{1,\dots,k\}$, and $\delta_{k+1}(B)$
correspond, respectively, to the number of times that
$\Gamma_\nabla$, $\eta_i\spV$, $i \in \{1,\dots,k\}$, and
$\eta\spV$ appear in $B$.  For each $l$, consider the following
sets
\begin{align*}
  \Br^l (\famX) = \setdef{B \; \text{bracket in} \, \famX}{\delta (B)
    = l} , \qquad \Br_l (\famX) = \setdef{B \; \text{bracket in} \,
    \famX}{B \in \Pc_l} ,
\end{align*}
where $\Pc_l$ denotes the set of homogeneous sections of
$\prol{E}$ of degree $l$.  The notion of \emph{primitive} bracket
will also be useful. Given a bracket $B$ in $\famX$, it is clear
that we can write $B=[B_1,B_2]$, with $B_i$ brackets in $\famX$.
In turn, we can also write $B_\alpha = [B_{\alpha1},B_{\alpha2}]$
for $\alpha=1,2$, and continue these decompositions until we end
up with elements belonging to $\famX$.  The collection of brackets
$B_1,B_2,B_{11},B_{12},\dots$ are called the \emph{components} of
$B$. The components of $B$ which do not admit further
decompositions are called \emph{irreducible}. A bracket $B$ is
called \emph{primitive} if all of its components are brackets in
$\Br_{-1}(\famX) \cup \Br_{0}(\famX) \cup \{ \Gamma_{\nabla} \}$.

Consider the set $\famX' = \{\Gamma_{\nabla} - \eta\spV,
\eta\spV_1,\dots,\eta\spV_k \}$. Clearly, the elements in
$\invclos{\famX'}$ are linear combinations of the elements in
$\invclos{\famX}$. In fact, for each bracket $B'$ of elements in
$\famX'$, let us define the subset $S(B') \subset \Br(\famX)$
formed by all possible brackets $B \in \Br (\famX)$ obtained by
replacing each occurrence of $\Gamma_{\nabla}-\eta\spV$ in $B'$ by
either $\Gamma_{\nabla}$ or $\eta\spV$. Then, one can prove by
induction (cf.~\cite{ADL:95a}) that
\begin{align}\label{eq:decomposition}
  B' = \sum_{B \in S(B')} (-1)^{\delta_{k+1} (B)} B  .
\end{align}
Reciprocally, given an element $B \in \Br (\famX)$, one can
determine the bracket $B'$ of elements in $\famX'$ such that $B
\in S(B')$ simply by substituting each occurrence of
$\Gamma_{\nabla}$ or $\eta\spV$ in $B$ by
$\Gamma_{\nabla}-\eta\spV$.  We denote this operation by $\psinv
(B) = B'$.  For each $k \in \natural$, define the following
families of sections in $E$,
\begin{align*}
  &\mathcal{C}^{(k)}_{\ver} (\eta;\eta_1,\dots,\eta_k) = \\
  & \setdef{\sigma \in \Sec{E}}{\sigma\spV = B'', B'' =
    \hspace*{-.25cm} \sum_{
      \begin{subarray}{c}
        \tilde{B} \in S(\psinv(B)) \\
        \cap\Br_{-1}(\famX) \cap \Br_{0}(\famX)
      \end{subarray}
    } \hspace*{-.75cm} (-1)^{\delta_{k+1}(\tilde{B})} \tilde{B} , \; B
    \in \Br^{2k-1}(\famX) \; \text{primitive}} ,
  \\
  &  \mathcal{C}^{(k)}_{\hor} (\eta;\eta_1,\dots,\eta_k) = \\
  & \setdef{\sigma \in \Sec{E}}{\sigma = \sigma_{B''}, B'' =
    \hspace*{-.25cm} \sum_{
    \begin{subarray}{c}
      \tilde{B} \in S(\psinv(B)) \\
      \cap\Br_{-1}(\famX) \cap \Br_{0}(\famX)
    \end{subarray}
  } \hspace*{-.75cm} (-1)^{\delta_{k+1}(\tilde{B})} \tilde{B}  , \;
  B \in \Br^{2k}(\famX) \; \text{primitive}}  .
\end{align*}
Consider
\begin{eqnarray*}
  \mathcal{C}_{\ver}
  (\eta;\eta_1,\dots,\eta_k) &=& \cup_{k \in
    \natural} \mathcal{C}^{(k)}_{\ver} (\eta;\eta_1,\dots,\eta_k),\\
  \mathcal{C}_{\hor}(\eta;\eta_1,\dots,\eta_k) &=& \cup_{k \in
    \natural} \mathcal{C}^{(k)}_{\hor} (\eta;\eta_1,\dots,\eta_k),
\end{eqnarray*}
and denote by $C_{\ver}(\eta;\eta_1,\dots,\eta_k)$ and $C_{\hor}
(\eta;\eta_1,\dots,\eta_k)$, respectively, the subbundles of the Lie
algebroid $E$ generated by the latter families.

Taking into account the previous discussion, we are now ready to
compute~$\invclos{\{\Gamma_{\nabla} -
  \eta\spV,\eta\spV_1,\dots,\eta\spV_k\}}$ for a mechanical control
system defined on a Lie algebroid.

\begin{proposition}(\cite{CoMa})
  \label{prop:decomposition-accessibility-distribution}
  Let $m\in M$. Then,
  \begin{align*}
    & \invclos{\{\Gamma_{\nabla} -
      \eta\spV,\eta\spV_1,\dots,\eta\spV_k\}} \cap
    \Ver_{0_m}(\prol{E}) = C_{\ver}(\eta;\eta_1,\dots,\eta_k)(m) \spV
    , \\
    & \invclos{\{\Gamma_{\nabla} -
      \eta\spV,\eta\spV_1,\dots,\eta\spV_k \}} \cap \Hor_{m}(\prol{E})
    = C_{\hor}(\eta;\eta_1,\dots,\eta_k)(m) .
  \end{align*}
\end{proposition}

\begin{remark}
  {\rm In the absence of potential terms, i.e., $\eta =0$, one has
    that
    \begin{align*}
      C_{\ver}(0;\eta_1,\dots,\eta_k) = \symclos{\{
        \eta_1,\dots,\eta_k\}} , \; C_{\hor}(0;\eta_1,\dots,\eta_k)
      = \invclos{\symclos{\{\eta_1,\dots,\eta_k\}}} ,
    \end{align*}
    where $\symclos{\{ \eta_1,\dots,\eta_k\}}$ denotes the
    distribution obtained by closing (the distribution defined by) $\{
    \eta_1,\dots,\eta_k\}$ under the symmetric product associated
    with~$\nabla$.  It is worth noticing that, in this case,
    $C_{\ver}(0;\eta_1,\dots,\eta_k) \subseteq
    C_{\hor}(0;\eta_1,\dots,\eta_k)$. This is not true in general.  }
\end{remark}

\subsection{Accessibility and controllability tests}

In this section we merge the notions introduced in
Section~\ref{se:notions} with the results obtained in
Section~\ref{se:structure} to give tests for accessibility and
controllability.

\begin{proposition}\cite{CoMa}\label{prop:accessibility}
  Let $m \in M$ and assume the Lie algebroid $E$ is locally transitive
  at $m$.  Then the mechanical control sys\-tem~\eqref{eq:CS} is
  \begin{itemize}
  \item locally base accessible from $m$ if $\,
    C_{\hor}(\eta;\eta_1,\dots,\eta_k)(m) + \ker \rho = E_{m}$,
  \item locally accessible from $m$ at zero if $\,
    C_{\hor}(\eta;\eta_1,\dots,\eta_k)(m) + \ker \rho = E_{m}$ and $\,
    C_{\ver}(\eta;\eta_1,\dots,\eta_k)(m)= E_{m}$.
  \end{itemize}
\end{proposition}

In order to state controllability tests, we need to introduce the
notions of good and bad symmetric products. We say that a
symmetric product $P$ in the sections $\{ \eta,\eta_1,\dots,\eta_k
\}$ is \emph{bad} if the number of occurrences of each $\eta_i$ in
$P$ is even. Otherwise, $P$ is \emph{good}. Accordingly,
$\symprod{\eta_i}{\eta_i}$ is bad and
$\symprod{\symprod{\eta}{\eta_j}}{\symprod{\eta_i}{\eta_i}}$ is
good. The following theorem gives sufficient conditions for local
controllability.

\begin{proposition}\cite{CoMa}\label{prop:controllability}
  Let $m \in M$.  The mechanical control sys\-tem~\eqref{eq:CS} is
  \begin{itemize}
  \item locally base controllable from $m$ if it is locally base
    accessible from $m$ and every bad symmetric product in $\{\eta,
    \eta_1,\dots,\eta_k \}$ evaluated at $m$ can be put as an
    $\real$-linear combination of good symmetric products of lower
    degree and elements of $\ker \rho$,
  \item locally controllable from $m$ at zero if it is locally
    accessible from $m$ at zero and every bad symmetric product in
    $\{\eta, \eta_1,\dots,\eta_k \}$ evaluated at $m$ can be put as an
    $\real$-linear combination of good symmetric products of lower
    degree.
  \end{itemize}
\end{proposition}

The corresponding tests for base accessibility and controllability
with regards to a manifold can be proved in a similar way.

\begin{proposition}\cite{CoMa}\label{prop:accessibility-with-regards-to-N}
  Let $\map{\psi}{M}{N}$ be an open map.  Let $m \in M$ and assume
  $\psi_* (\rho (E_m)) = T_{\psi(m)}N$.  Then the mechanical control
  sys\-tem~\eqref{eq:CS} is
  \begin{itemize}
  \item locally base accessible from $m$ with regards to $N$ if $\,
    C_{\hor}(\eta;\eta_1,\dots,\eta_k)(m) + \rho^{-1} (\ker \psi_*) =
    E_{m}$,
  \item locally base controllable from $m$ with regards to $N$ if the
    system is locally base accessible from $m$ with regards to $N$ and
    every bad symmetric product in $\{ \eta,\eta_1,\dots,\eta_k \}$
    evaluated at $m$ can be put as an $\real$-linear combination of
    good symmetric products of lower degree and elements of $\rho^{-1}
    (\ker \psi_*)$.
  \end{itemize}
\end{proposition}

\section{Discrete Mechanics on Lie groupoids}\label{discrete}

In this section, we discuss discrete Lagrangian Mechanics on a Lie
groupoid $G\rightrightarrows M$. Instead of the usual Euler-Lagrange
equations (\ref{free-forces}) for a Lie algebroid $\tau: E\to M$
equipped with a Lagrange function $L: E\to \R$, we obtain a set of
difference equations called {\sl Discrete Euler-Lagrange equations}
for a discrete Lagrangian $L_d: G\to \R$ \cite{groupoid}. When the Lie
algebroid is precisely $E=E_G$ and $L_d$ is a suitable approximation
of the continuous Lagrangian $L: E_G\to \R$, then we will obtain a
geometric integrator for the Euler-Lagrange equations. In the next
subsections we will carefully analyze this construction and its
geometric properties.

\subsection{Lie algebroid structure on the vector bundle
  $\pi^\tau:\Gprol{E_G}\maparrow G$}

Let $G \rightrightarrows M$ be a Lie groupoid with structural maps
\[
\alpha, \beta: G \to M, \; \; \epsilon: M \to G, \; \; i: G \to G, \;
\; m: G_{2} \to G.
\]
We know that $\Gprol{E_G}\rightrightarrows E_G$ is a Lie groupoid. The
following theorem shows that the vector bundle
$\pi^\tau:\Gprol{E_G}\cong V\beta \oplus_{G} V\alpha \to G$ is
equipped with a natural structure of Lie algebroid.

\begin{theorem}[See Theorem 3.3 in   \cite{groupoid}] \label{t3.1}
  The vector bundle $\pi^\tau:\Gprol{E_G}\cong V\beta \oplus_{G}
  V\alpha \to G$ admits a Lie algebroid structure, where the anchor
  map is given by
  \begin{equation} \label{e1'}
    \rho^{\Gprol{E_G}}(X_{g}, Y_{g}) =
    X_{g} + Y_{g}, \; \; \mbox{ for } (X_{g}, Y_{g}) \in V_{g}\beta
    \oplus V_{g}\alpha,
  \end{equation}
  and the Lie bracket $\lcf\cdot , \cdot\rcf^{\Gprol{E_G}}$ on the
  space $\Sec{\pi^\tau}$ is characterized by the following
  relation
  \begin{equation}\label{e1''}
    \lcf (\rvec{X}, \lvec{Y}), (\rvec{X'}, \lvec{Y'}) \rcf
    ^{\Gprol{E_G}} =\Bigl(-\overrightarrow{\lcf X, X' \rcf},
    \overleftarrow{\lcf Y, Y' \rcf}\Bigr),
  \end{equation}
  for $X, Y, X', Y' \in \Sec{\tau}$.
\end{theorem}
We also remark that, if we denote by $(G,\alpha)$ the fibration
$\map{\alpha}{G}{M}$ and by $(G,\beta)$ the fibration
$\map{\beta}{G}{M}$, then it is not difficult to prove that the Lie
algebroid prolongations $\prol[E_G]{(G,\alpha)}\maparrow G$ and
$\prol[E_G]{(G,\beta)}\maparrow G$ are isomorphic, as Lie algebroids,
and both are isomorphic to the Lie algebroid $V\beta \oplus_{G}
V\alpha\maparrow G$ and hence to $\Gprol{E_G}\maparrow G$
(see~\cite{groupoid} for the details).

The following diagram shows both structures of $\Gprol{E_G}$,
\[
\xymatrix{%
  &\Gprol{E_G}\ar@{=>}[rr]^{\ \ \ \alpha^{\tau}}_{\ \ \
    \beta^{\tau}}\ar[rd]^{\rho^{\Gprol{E_G}}}
  \ar[dd]^{\pi^\tau}&&E_G\ar[rd]^{\rho}\ar'[d][dd]^{\tau}\\
  &&TG\ar@{=>}[rr]^(.4){T\alpha}_(.4){T\beta}\ar[ld]_(.6){\tau_G}&&TM\ar[ld]^{\tau_M}\\
  &G\ar@{=>}[rr]^\alpha_\beta&&M }
\]
where the vertical maps are morphisms of Lie groupoids and the
horizontal maps are morphisms of Lie algebroids.

Given a section $X$ of $E_G\to M$, we define the sections $X^{(1,
  0)}$, $X^{(0,1)}$ (the $\beta$ and $\alpha$- lifts) and $X^{(1, 1)}$
(the complete lift) of $X$ to $\pi^\tau: \Gprol{E_G}\to G$ as follows:
\[
X^{(1, 0)}(g)=(\rvec{X}(g), 0_g), \quad X^{(0, 1)}(g)=(0_g,
\lvec{X}(g)) \quand X^{(1, 1)}(g)=(-\rvec{X}(g), \lvec{X}(g))
\]
We can easily see that
\begin{equation}\label{e1''*}
\begin{array}{l}
\lcf X^{(1, 0)}, Y^{(1, 0)}\rcf ^{\Gprol{E_G}} = -\lcf X,
Y \rcf^{(1,0)} \\
 \lcf X^{(0, 1)}, Y^{(0, 1)}\rcf ^{\Gprol{E_G}} =\hphantom{-} \lcf X, Y \rcf^{(0,1)}
\end{array}
 \hbox{  and  } \ \lcf X^{(0,
1)}, Y^{(1, 0)}\rcf ^{\Gprol{E_G}} = 0
\end{equation}
and, as a consequence,
\begin{equation}\label{e2''}
\begin{array}{l}
\lcf X^{(1, 1)}, Y^{(1, 0)}\rcf ^{\Gprol{E_G}} = \lcf X,
Y \rcf^{(1,0)}  \\
\lcf X^{(1, 1)}, Y^{(0, 1)}\rcf ^{\Gprol{E_G}} = \lcf X, Y \rcf
^{(0,1)}
\end{array}
 \hbox{  and  }\ \lcf X^{(1,
1)}, Y^{(1, 1)}\rcf ^{\Gprol{E_G}} = \lcf X, Y \rcf^{(1,1)}.
\end{equation}

\subsection{Discrete Variational Mechanics on Lie groupoids}

Discrete Lagrangian systems on Lie groupoids have a variational
origin, as we explain next. A \emph{discrete Lagrangian system}
consists of a Lie groupoid $G\rightrightarrows M$ (the \emph{discrete
  space}) and a \emph{discrete Lagrangian} $L_d: G \to \R$.

\paragraph{\bf Discrete Euler-Lagrange equations} For $g\in G$ fixed,
we consider the set of \emph{admissible sequences}:
\[
{\mathcal C}^N_{g}=\set{(g_1, \ldots, g_N)\in G^N}{(g_k, g_{k+1})\in
  G_2 \hbox{ for } k=1,\ldots, N-1 \hbox{ and } g_1 \ldots g_n=g }.
\]
It is easy to show that we may identify the tangent space to
${\mathcal C}^N_g$ with
\[
T_{(g_1, \ldots, g_N)}{\mathcal C}^N_g\equiv\set{(v_1, \ldots,
  v_{N-1})}{v_k\in (E_G)_{x_k}\hbox{ and } x_k=\beta(g_k), 1\leq k\leq
  N-1}.
\]
An element of $T_{(g_1, \ldots, g_N)}{\mathcal C}^N_g$ is called an
\emph{ infinitesimal variation}. Now, we define the \emph{discrete
  action sum} associated to the discrete Lagrangian $L_d: G\to \R$ by
\[
{\mathcal S} L_d( (g_1, \ldots, g_N)= \sum_{k=1}^{N} L_d(g_k).
\]
Hamilton's principle requires that this discrete action sum be
stationary with respect to all the infinitesimal variations. This
requirement gives the following alternative expressions for the
\emph{discrete Euler-Lagrange equations} (see \cite{groupoid}):
\begin{eqnarray}
  \lvec{X}\big({g_k})(L_d)-\rvec{X}\big({g_{k+1}})(L_d)=0,
  \label{discretee}
\end{eqnarray}
or
\begin{eqnarray*}
  \pai{d L_d}{X^{(0,1)}}(g_k)-\pai{d
    L_d}{X^{(1,0)}}(g_{k+1})= 0,
\end{eqnarray*}
for all sections $X$ of $\tau:E_G\to M$. Here, $d$ denotes the
differential of the Lie algebroid $\pi^\tau: \Gprol{E_G}\equiv
V\beta\oplus_G V\alpha\to G$. Alternatively, we may rewrite the
Discrete Euler-Lagrange equations as
\[
\d\left[L_d\circ l_{g_k}+L_d\circ r_{g_{k+1}}\circ
i\right](\epsilon (x_k))_{\big|(E_G)_{x_k}}=0,
\]
where $\beta(g_k)=\alpha(g_{k+1})=x_k$, and where $\d$ denotes the
standard differential on $G$, that is, the differential of the Lie
algebroid $\tau_G: TG\to G$.

Thus, we may define the \emph{discrete Euler-Lagrange operator}:
\[
D_{\hbox{\footnotesize DEL}}L_d: G_2\to E^*_G\; ,
\]
where $E^*_G$ is the dual of $E_G$. This operator is given by
\[
D_{\hbox{\footnotesize DEL}}{L_d}(g, h)=  \d\left[L_d\circ
l_{g}+L_d\circ r_{h}\circ i\right](\epsilon (x))_{\big|(E_G)_{x_k}}
\]
with $\beta(g)=\alpha(h)=x$.

\paragraph{\textbf{Discrete Poincar\'e-Cartan sections}} Consider the
Lie algebroid $\pi^\tau:\Gprol{E_G}\cong V\beta\oplus_G V\alpha
\to G$, and define the \emph{Poincar\'e-Cartan 1-sections }
$\Theta_{L_d}^-, \Theta_{L_d}^+\in \Sec{ (\pi^\tau)^*}$ as follows
\begin{equation}\label{5.16'}
\Theta_{L_d}^-(g)(X_g, Y_g)= -X_g(L_d),\;\;\;\;\;
\Theta_{L_d}^+(g)(X_g, Y_g)= Y_g(L_d),
\end{equation}
 for each $g\in G$ and $(X_g, Y_g)\in V_g\beta\oplus
V_g\alpha$.

Since $dL_d=\Theta_{L_d}^+ - \Theta_{L_d}^- $ and so, using $d^2=0,$
it follows that $d\Theta_{L_d}^+=d\Theta_{L_d}^-$. This means that
there exists a unique 2-section
$\Omega_{L_d}=-d\Theta_{L_d}^+=-d\Theta_{L_d}^-$, that will be called
the \emph{Poincar\'e-Cartan} 2-section. This 2-section will be
important to study the symplectic character of the discrete
Euler-Lagrange equations.

If $\{X_i\}$ is a local basis of ${\rm Sec}(\tau)$ then
$\{X_i^{(1,0)},X_i^{(0,1)}\}$ is a local basis of $\Sec{\pi^\tau}.$
Moreover, if $\{(X^i)^{(1,0)}, (X^i)^{(0,1)}\}$ is the dual basis of
$\{X_i^{(1,0)},X_i^{(0,1)}\}$, it follows that
\[
\Theta_{L_d}^-=-\rvec{X}_i(L_d)(X^i)^{(1,0)},\;\;\;
\Theta_{L_d}^+=\lvec{X}_i(L_d)(X^i)^{(0,1)},
\]
\[
\Omega_{L_d}=-\rvec{X}_i(\lvec{X}_jL_d)(X^i)^{(1,0)}\wedge
(X^j)^{(0,1)}.
\]

\paragraph{\bf Discrete Lagrangian evolution operator} Let $\xi: G\to
G$ be a smooth map such that:
\begin{enumerate}
\item[-] $\hbox{graph}(\xi)\subseteq G_2$, that is, $(g, \xi(g))\in
  G_2$, for all $g\in G$ ($\xi$ is a \emph{second order operator}).
\item[-] $(g, \xi(g))$ is a solution of the discrete Euler-Lagrange
  equations, for all $g\in G$, that is,
  $(D_{\hbox{DEL}}L_d)(g,\xi(g))=0,$ for all $g\in G.$
\end{enumerate}
In such  case
\begin{equation}\label{5.22'}
  \lvec{X}(g)(L_d)-\rvec{X}(\xi(g))(L_d)=0
\end{equation}
for every section $X$ of $E_G$ and every $g\in G.$ The map $\xi: G\to
G$ is called a \emph{discrete flow} or a \emph{discrete Lagrangian
  evolution operator for $L_d$}.

Now, let $\xi:G\to G$ be a second order operator. Then, the
prolongation $\prol[]{\xi}: V\beta\oplus_G V\alpha\to V\beta\oplus_G
V\alpha$ of $\xi$ is the Lie algebroid morphism over $\xi: G\to G$
defined as follows (see \cite{groupoid}):
\begin{equation}\label{poi}
  \prol[]{\xi}[g](X_g, Y_g)=((T_g(r_{g\xi(g)}\circ i))(Y_g),
  (T_g\xi)(X_g)+(T_g\xi)(Y_g)-T_g(r_{g\xi(g)}\circ i)(Y_g)) ,
\end{equation}
for all $(X_g, Y_g)\in V_g\beta\oplus V_g\alpha$. Moreover, from
(\ref{linv}), (\ref{rinv}) and (\ref{poi}),
we obtain that
\begin{equation}\label{zaa}
  \prol[]{\xi}[g](\rvec{X}(g), \lvec{Y}(g))=(-\rvec{Y}(\xi(g)),
  (T_g\xi)(\rvec{X}(g)+\lvec{Y}(g)) +\rvec{Y}(\xi(g))) ,
\end{equation}
for all $X, Y$ sections of $E_G$.

Using (\ref{poi}), one may prove that (see \cite{groupoid}):
\begin{enumerate}
\item The map $\xi$ is a discrete Lagrangian evolution operator for
  $L_d$ if and only if
  $(\prol[]{\xi},\xi)^*\Theta_{L_d}^-=\Theta_{L_d}^+$.
\item The map $\xi$ is a discrete Lagrangian evolution operator for
  $L_d$ if and only if
  $(\prol[]{\xi},\xi)^*\Theta_{L_d}^--\Theta_{L_d}^-=dL_d$.
\item If $\xi$ is discrete Lagrangian evolution operator then
  $(\prol[]{\xi},\xi)^*\Omega_{L_d}=\Omega_{L_d}$.
\end{enumerate}

\paragraph{\bf Discrete Legendre transformations} Given a Lagrangian
$L_d: G\to \R$ we define the \emph{discrete Legendre transformations}
$\F^{-}L_d: G\to E^*_G$ and $\F^{+}L_d: G\to E^*_G$ by
\begin{align*}
  (\F^{-}L_d)(h)(v_{\epsilon(\alpha(h))})&= -v_{\epsilon(\alpha(h))}(L_d\circ
  r_h\circ i),\;\;\; \mbox{ for }
  v_{\epsilon(\alpha(h))}\in (E_G)_{\alpha(h)},\\
  (\F^{+}L)(g)(v_{\epsilon(\beta(g))})&= v_{\epsilon(\beta(g))}(L_d\circ
  l_g), \mbox{ for } v_{\epsilon(\beta(g))}\in
  (E_G)_{\beta(g)}.
\end{align*}
Now, we introduce the prolongations $\prol[]{\F^{-}L_d}:
\Gprol{E_G}\equiv V\beta\oplus_G V\alpha\to \prol[E_G]{E^*_G}$ and
$\prol[]{\F^{+}L_d}: \Gprol{E_G}\equiv V\beta\oplus_G V\alpha\to
\prol[E_G]{E^*_G}$ by
\begin{align*}
  \prol[]{\F^{-}L_d}[h](X_h, Y_h) &= (T_h(i\circ
  r_{h^{-1}})(X_h),  (T_h\F^{-}L)(X_h)+(T_h\F^{-}L)(Y_h)),\\
  \prol[]{\F^{+}L_d}[h](X_h, Y_h) &= (T_h l_{h^{-1}}(Y_h),
  (T_h\F^{+}L)(X_h)+(T_h\F^{+}L)(Y_h)),
\end{align*}
for all $h\in G$ and $(X_h, Y_h)\in V_h\beta\oplus V_h\alpha$. We
observe that the discrete Poincar\'e-Cartan 1-sections and
2-section are related to the canonical Liouville section of
$(\prol[E_G]{E^*_G})^*\to E^*_G$ and the canonical symplectic
section of $\wedge^2(\prol[E_G]{E^*_G})^*\to E^*_G$ by pull-back
under the discrete Legendre transformations, that is,
\begin{eqnarray*}
  (\prol[]{\F^{-}L_d}, \F^-L_d)^*\Theta_{E_G}= \Theta^-_{L_d},&&
  (\prol[]{\F^{+}L_d}, \F^+L_d)^*\Theta_{E_G}= \Theta^+_{L_d},\\
  (\prol[]{\F^{-}L_d}, \F^-L_d)^*\Omega_{E_G}= \Omega_{L_d}, &&
  (\prol[]{\F^{+}L_d}, \F^+L_d)^*\Omega_{E_G}= \Omega_{L_d} .
\end{eqnarray*}

\paragraph{\bf Discrete regular Lagrangians} A discrete Lagrangian
$L_d:G\to \R$ is said to be \emph{regular} if the set of solutions of
the discrete Euler-Lagrange equations is locally the graph of a
diffeomorphism, that is, there exists locally a unique discrete
Lagrangian evolution operator $\xi_{L_d}: G\to G$ for $L_d$.  In such
a case, $\xi_{L_d}$ is called the discrete Euler-Lagrange evolution
operator. In \cite{groupoid} (see Theorem 4.13 in \cite{groupoid}), we
obtained some necessary and sufficient conditions for a discrete
Lagrangian on a Lie groupoid $G$ to be regular that we summarize as
follows:
\begin{eqnarray*}
  L_d \hbox{ is regular} &\Longleftrightarrow&  \hbox{The Legendre
    transformation } \F^+L_d  \hbox{ is a local
    diffeomorphism}\\
  &\Longleftrightarrow&  \hbox{The Legendre transformation } \F^-L_d
  \hbox{ is a local
    diffeomorphism}\\
  &\Longleftrightarrow& \hbox{The Poincar\'{e}-Cartan $2$-section
    $\Omega_{L_d}$ is symplectic}\\
  && \hbox{on the Lie algebroid $\Gprol{E_G}\equiv
    V\beta\oplus_G V\alpha\to G$.}
\end{eqnarray*}
Locally, we deduce that $L_d$ is regular if and only if for every
local basis $\{X_i\}$ of ${\rm Sec}(\tau)$ the local matrix
$(\rvec{X}_i(\lvec{X}_jL_d))$ is regular.

\paragraph{\bf Discrete Hamiltonian evolution operator} If $L_d: G\to
\R$ is a regular Lagrangian, then pushing forward to $E^*_G$ with the
discrete Legendre transformations, we obtain the \emph{discrete
  Hamiltonian evolution operator}, $\tilde{\xi}_{L_d}: E^*_G\to E^*_G$
which is given by
\begin{equation}\label{dheo}
\tilde{\xi}_{L_d}=\F^{\pm}L_d\circ \xi_{L_d}\circ
(\F^{\pm}L_d)^{-1}\;.
\end{equation}
Defining the prolongation
$\prol[]{\tilde{\xi}_{L_d}}:\prol[E_G]{E^*_G}\to \prol[E_G]{E^*_G}$ of
$\tilde{\xi}_{L_d}$ by
$$\prol[]{\tilde{\xi}_{L_d}}=\prol[]{\F^{\pm}{L_d}}\circ
\prol[]{\xi_{L_d}}\circ (\prol[]{\F^{\pm}{L_d}})^{-1},$$
we deduce
that (see \cite{groupoid}):
\[
(\prol[]{\tilde{\xi}_{L_d}}, \tilde{\xi}_{L_d})^*\Theta_{E_G}=
\Theta_{E_G} + d(L_d \circ (\F^-{L_d})^{-1}), \makebox[.4cm]{}
(\prol[]{\tilde{\xi}_{L_d}}, \tilde{\xi}_{L_d})^*\Omega_{E_G}=
\Omega_{E_G}.
\]

\paragraph{\bf Noether's theorem} In Discrete Mechanics is also
possible to relate invariance of the discrete Lagrangian under some
transformation group with the existence of constants of the motion. In
fact, we will say that a section $X$ of $E_G$ is a \emph{Noether's
  symmetry of the Lagrangian} $L_d$ if there exists a function
$f\in\cinfty{M}$ such that
\[
dL_d(X^{(1,1)})=\beta^*f-\alpha^*f.
\]
In the particular case when $dL_d(X^{(1,1)})=-\rvec{X}L_d +
\lvec{X}L_d= 0$, we  will say that  $X$ is an \emph{infinitesimal
symmetry} of the discrete Lagrangian $L_d$.

If $L_d:G\to \R$ is a regular discrete Lagrangian, by a \emph{constant
  of the motion } we mean a function $F$ invariant under the discrete
Euler-Lagrange evolution operator $\xi_{L_d}$, that is,
$F\circ\xi_{L_d}=F$. Then, we have the following result.

\begin{theorem}[Discrete Noether's theorem] \cite{groupoid} If $X$ is
  a Noether symmetry of a discrete Lagrangian $L_d$, then the function
  $F={\Theta^-_{L_d}}(X^{(1,1)})-\alpha^*f$ is a \emph{constant of the
    motion } for the discrete dynamics defined by $L_d$.
\end{theorem}

\section{Classical Field Theory on Lie algebroids}

In this section, we study Classical Field Theories on Lie algebroids.
We consider a fiber bundle $\map{\nu}{M}{N}$, a Lie algebroid
structure on a vector bundle $\map{\prEM}{E}{M}$ and a surjective
morphism of Lie algebroids $\map{\prEF}{E}{TN}$ over~$\prMN$.  The
physical interpretation of the above data is as follows: we will
consider a field theory in which the fields are the sections of the
bundle $\prMN$ and the partial derivatives of the fields are
parameterized by linear sections of $\prEF$.

We will find the equations for the extremals of a variational problem
which roughly speaking is the following: given a Lagrangian function
$L$ defined on the set of sections of $\prEF$, and a volume form
$\omega$ on the manifold $N$, we look for those morphisms of Lie
algebroids which are critical points of the action functional $$
\CMcal{S}(\Phi)=\int_N L(\Phi)\,\omega.  $$
This is a constrained
variational problem, because we are restricting the fields $\Phi$ to
be morphisms of Lie algebroids, which is a condition on the
derivatives of~$\Phi$.

\subsection{Jets}
\label{jetoids}

We consider two vector bundles $\map{\prEM}{E}{M}$ and
$\map{\prFN}{F}{N}$ and a surjective vector bundle map
$\map{\prEF}{E}{F}$ over the map $\map{\prMN}{M}{N}$. Moreover, we
will assume that $\map{\prMN}{M}{N}$ is a smooth fiber bundle. We will
denote by $K\to M$ the kernel of the map $\prEF$, which is a vector
bundle over~$M$. Given a point $m\in M$, if we denote $n=\prMN(m)$, we
have the following exact sequence $0\to K_m \to E_m \to F_n \to 0$,
and we can consider the set $\Jpi[m]$ of splittings $\phi$ of such
sequence. More concretely, we define the following sets
$\Lpi[m]=\set{\map{w}{F_n}{E_m}}{\text{$w$ is linear}}$,
$\Jpi[m]=\set{\phi\in\Lpi[m]}{\prEF\circ\phi=\id_{F_n}}$ and
$\Vpi[m]=\set{\psi\in\Lpi[m]}{\prEF\circ\psi=0}$.  Therefore $\Lpi[m]$
is a vector space, $\Vpi[m]$ is a vector subspace of $\Lpi[m]$ and
$\Jpi[m]$ is an affine subspace of $\Lpi[m]$ modeled on the vector
space $\Vpi[m]$. By taking the union, $\Lpi=\cup_{m\in M}\Lpi[m]$,
$\Jpi=\cup_{m\in M}\Jpi[m]$ and $\Vpi=\cup_{m\in M}\Vpi[m]$, we get
the vector bundle $\map{\prLM}{\Lpi}{M}$ and the affine subbundle
$\map{\prJM}{\Jpi}{M}$ modeled on the vector bundle
$\map{\prVM}{\Vpi}{M}$. We will also consider the projection
$\map{\prJN}{\Jpi}{N}$ defined by composition $\prJN=\prMN\circ\prJM$.
An element of $\Jpi[m]$ will be simply called a \emph{jet} at the
point $m\in M$ and accordingly the bundle $\Jpi$ is said to be the
first \emph{jet bundle} of $\prEF$.

\smallskip

Notice that the standard case~\cite{Saunders} is recovered when we
have a bundle $\map{\prMN}{M}{N}$ and one considers the standard Lie
algebroids $E=TM\to M$ and $F=TN\to N$ together with the differential
of the projection $\map{\pi=T\prMN}{TM}{TN}$. With the standard
notations, we have that $J^1\nu\equiv\mathcal{J}(T\nu)$.

\medskip

Local coordinates on $\Jpi$ are given as follows. We consider local
coordinates $(x^i)$ on $N$ and $(x^i,u^A)$ on $M$ adapted to the
projection $\prMN$. We also consider local basis of sections
$\{\bar{e}_a\}$ of $F$ and $\{e_a,e_\alpha\}$ of $E$ adapted to the
projection $\prEF$, that is $\prEF\circ e_a=\bar{e}_a\circ\prMN$ and
$\prEF\circ e_\alpha=0$. In this way $\{e_\alpha\}$ is a base of
sections of $K$. An element $w$ in $\Lpi[m]$ is of the form
$w=(w_a^be_b+w_a^\alpha e_\alpha)\otimes \bar{e}^a$, and it is in
$\Jpi[m]$ if and only if $w_a^b=\delta_a^b$, i.e., an element $\phi$
in $\Jpi$ is of the form $\phi=(e_a+\phi^\alpha_a
e_\alpha)\otimes\bar{e}^a$. If we set
$y^\alpha_a(\phi)=\phi^\alpha_a$, we have adapted local coordinates
$(x^i,u^A,y^\alpha_a)$ on $\Jpi$. Similarly, an element
$\psi\in\Vpi[m]$ is of the form $\psi=\psi^\alpha_a
e_\alpha\otimes\bar{e}^a$. If we set $y^\alpha_a(\psi)=\psi^\alpha_a$,
we have adapted local coordinates $(x^i,u^A,y^\alpha_a)$ on $\Vpi$. As
usual, we use the same name for the coordinates in an affine bundle
and in the associated vector bundle.

An element $z\in\dLpi[m]$ defines an affine function $\hat{z}$ on
$\Jpi[m]$ by contraction $\hat{z}(\phi)=\pai{z}{\phi}$ where
$\pai{\cdot}{\cdot}$ is the pairing
$\pai{z}{\phi}=\tr(z\circ\phi)=\tr(\phi\circ z)$. Therefore, a section
$\theta$ of $\dLpi$ defines a fiberwise affine function $\hat{\theta}$
on $\Jpi$, $ \hat{\theta}(\phi)=\pai{\theta_{\prJM(\phi)}}{\phi}=
\tr(\theta_{\prJM(\phi)}\circ\phi). $ In local coordinates, a section
of $\dLpi$ is of the form
$\theta=(\theta^a_b(x)e^b+\theta^a_\alpha(x)e^\alpha)\otimes\bar{e}_a$,
and the affine function defined by $\theta$ is
$\hat{\theta}=\theta^a_a(x)+\theta^a_\alpha(x)y^\alpha_a$.

\paragraph{\bf Anchor} Consider now anchored structures on the bundles
$E$ and $F$, that is, we have two vector bundle maps
$\map{\rho_F}{F}{TN}$ and $\map{\rho_E}{E}{TM}$ over the identity in
$N$ and $M$ respectively. We will assume that the map $\prEF$ is
admissible, that is $\rho_F\circ\prEF=T\prMN\circ\rho_E$.  Therefore
we have $$
\rho_F(\bar{e}_a)=\rho^i_a\pd{}{x^i} \qquand
\left\{\begin{array}{l}
    \rho_E(e_a)=\rho^i_a\dpd{}{x^i}+\rho^A_a\dpd{}{u^A},\\[5pt]
    \rho_E(e_\alpha)=\rho^A_\alpha\dpd{}{u^A},
\end{array}\right.
$$
with $\rho^i_a=\rho^i_a(x)$, $\rho^A_a=\rho^A_a(x,u)$ and
$\rho^A_\alpha=\rho^A_\alpha(x,u)$.

The anchor allows us to define the concept of total derivative of a
function with respect to a section. Given a section
$\sigma\in\sec{F}$, the total derivative of a function
$f\in\cinfty{M}$ with respect to $\sigma$ is the function
$\widehat{df\otimes\sigma}$, i.e., the affine function associated to
$df \otimes \sigma \in \sec{\dLpi}$. In particular, the total
derivative with respect to an element $\bar{e}_a$ of the local basis
of sections of $F$, will be denoted by $\p{f}_{|a}$. In this way, if
$\sigma=\sigma^a\bar{e}_a$ then $
\widehat{df\otimes\sigma}=\p{f}_{|a}\sigma^a, $ where the coordinate
expression of $\p{f}_{|a}$ is $$
\p{f}_{|a}=\rho^i_a\pd{f}{x^i}+(\rho^A_a+\rho^A_\alpha
y^\alpha_a)\pd{f}{u^A}.  $$
Notice that, for a function $f$ in the
base $N$, we have that $\p{f}_{|a}=\rho^i_a\pd{f}{x^i}$ are just the
components of $df$ in the basis $\{\bar{e}_a\}$.

\paragraph{\bf Bracket} Finally, let us assume that we have Lie
algebroid structures on $\map{\prFN}{F}{N}$ and on
$\map{\prEM}{E}{M}$, and that the projection $\prEF$ is a morphism of
Lie algebroids. This condition implies the vanishing of some structure
functions.

We have the following expressions for the brackets of elements in the
basis of sections
$$
\lcf\bar{e}_a,\bar{e}_b\rcf=C^c_{ab}\bar{e}_c \qquand
\left\{\begin{aligned}
    &\lcf e_a,e_b\rcf=C^\gamma_{ab}e_\gamma+C^c_{ab}e_c\\
    &\lcf e_a,e_\beta\lcf=C^\gamma_{a\beta}e_\gamma \\
    &\lcf e_\alpha,e_\beta\rcf=C^\gamma_{\alpha\beta}e_\gamma,
 \end{aligned}\right.
$$
where $C^a_{bc}=C^a_{bc}(x)$ is a basic function.

The structure functions can be conveniently combined as terms of some
affine function as follows
\[
Z^\alpha_{a\gamma}=C^\alpha_{a\gamma}+C^\alpha_{\beta\gamma}y^\beta_a
\qquand Z^\alpha_{ac}=C^\alpha_{ac}+C^\alpha_{\beta c}y^\beta_a.
\]
In particular, such functions will appear in the Euler-Lagrange
equations of the variational problem.

\subsection{Morphisms and admissible maps}
\label{morphisms}

By a section of $\prEF$ we mean a vector bundle map $\Phi$ such
that $\prEF\circ\Phi=\id_F$, (i.e., we consider only linear
sections of $\prEF$ (see also~\cite{GrGrUr})). It follows that the
base map $\map{\phi}{N}{M}$ is a section of $\prMN$, i.e.,
$\prMN\circ\phi=\id_N$. The set of sections of $\prEF$ will be
denoted by $\sec{\prEF}$. The set of those sections of $\prEF$
which are a morphism of Lie algebroids will be denoted by
$\Mor{\prEF}$. We will find local conditions for
$\Phi\in\sec{\prEF}$ to be an admissible map between anchored
vector bundles and also local conditions for $\Phi$ to be a
morphism of Lie algebroids.

Taking adapted local coordinates $(x^i,u^A)$ on $M$, the map $\phi$
has the expression $\phi(x^i)=(x^i,u^A(x))$. If we moreover take an
adapted basis $\{e_a,e_\alpha\}$ of local sections of $E$, then the
expression of $\Phi$ is given by
$\Phi(\bar{e}_a)=e_a+y^\alpha_a(x)e_\alpha$, so that the map $\Phi$ is
determined by the functions $\bigl(u^A(x),y^\alpha_a(x)\bigr)$ locally
defined on $N$. The action on the dual basis is $\Phi\pb
e^a=\bar{e}^a$, and $\Phi\pb e^\alpha=y^\alpha_a(x) \bar{e}^a$, and
for the coordinate functions $\Phi\pb x^i=x^i$ and $\Phi\pb
u^A=u^A(x)$.

The admissibility condition reads $\Phi\pb(df)=d(\Phi\pb f)$ for every
function $f\in\cinfty{M}$. Taking $f=x^i$ we get an identity, while
taking $f=u^A$ we get the condition
\[
\rho^i_a\pd{u^A}{x^i}=\rho^A_a+\rho^A_\alpha y^\alpha_a.
\]

In addition to the admissibility condition, the morphism condition
reads $\Phi\pb d\theta=d(\Phi\pb\theta)$ for every section $\theta$ of
$E^*$. For $\theta=e^a$ we get an identity, while for
$\theta=e^\alpha$ we find the
\[
\rho^i_b\pd{y^\alpha_c}{x^i}-\rho^i_c\pd{y^\alpha_b}{x^i} -y^\alpha_a
C^a_{bc} + C^\alpha_{\beta\gamma}y^\beta_by^\gamma_c
+C^\alpha_{b\gamma}y^\gamma_c-C^\alpha_{c\gamma}y^\gamma_b
+C^\alpha_{bc}=0.
\]

\subsection{Variational Calculus}
\label{variational}

In what follows in this paper we consider the case where the Lie
algebroid $F$ is the tangent bundle $F=TN$ with $\rho_F=\id_{TN}$ and
$[\cdot,\cdot]$ the usual Lie bracket of vector fields on $N$.  The Lie
algebroid $E$ remains a general Lie algebroid. Moreover, for local
expressions on $F$, the local basis of sections of $F$ which we will
consider is a basis of coordinate vector fields $\bar{e}_i=\pd{}{x^i}$,
so that $\rho^i_a=\delta^i_a$ and $C^a_{bc}=0$.

\paragraph{\bf Variational problem} Given a Lagrangian function
$L\in\cinfty{\Jpi}$ and a volume form $\omega\in\ext[r]{(TN)}$, where
$r=\operatorname{dim}(N)$, we consider the following variational
problem: find the critical points of the action functional $
\CMcal{S}(\Phi)=\int_N L(\Phi)\,\omega $ defined on the set of
sections of $\prEF$ which are moreover morphisms of Lie algebroids,
that is, defined on the set $\Mor{\prEF}$. Here by $L(\Phi)$ we mean
the function $n\mapsto L(\Phi_n)$, where $\Phi_n\in\Jpi$ is the
restriction of $\Phi$ the fiber $F_n=T_nN$.

It is important to notice that the above variational problem is a
constrained problem, not only because the condition
$\prEF\circ\Phi=\id_F$, which can be easily solved, but because of the
condition of $\Phi$ being a morphism of Lie algebroids, which is a
condition on the derivatives of $\Phi$. Taking coordinates on $N$ such
that the volume form is $\omega=dx^1\wedge \cdots\wedge dx^r$, the
problem is to find the critical points of
\[
\int_N L(x^i,u^A(x),y^\alpha_a(x))\,dx^1\wedge \cdots\wedge dx^r ,
\]
subject to the constraints
\[
\pd{u^A}{x^a}=\rho^A_a+\rho^A_\alpha y^\alpha_a \quand
\pd{y^\alpha_c}{x^b}-\pd{y^\alpha_b}{x^c}+C^\alpha_{b\gamma}y^\gamma_c
-C^\alpha_{c\gamma}y^\gamma_b
+C^\alpha_{\beta\gamma}y^\beta_by^\gamma_c +C^\alpha_{bc}=0.
\]

The first method one can try to solve the problem is to use Lagrange
multipliers. Nevertheless, one has no warranties that all solutions to
this problem are normal (i.e., not strictly abnormal).  In fact, in
simple cases such as the problem of a heavy top~\cite{M}, one can
easily see that there will be strictly abnormal solutions.  Therefore
we take another approach, which consists of finding explicitly finite
variations of a solution, that is, defining a curve in $\Mor{\prEF}$
starting at the given solution.

\paragraph{\bf Variations and infinitesimal variations} In order to
find admissible variations, we consider sections of $E$ and the
associated flow. With the help of this flow we can transform morphisms
of Lie algebroids into morphisms of Lie algebroids.

\paragraph{\bf Flow defined by a section} We recall that every section
of a Lie algebroid has an associated local flow composed of morphisms
of Lie algebroids~\cite{CFTLAVA,MaXu}. More explicitly, given a
section $\sigma$ of a Lie algebroid $E$, there exists a local flow
$\map{\Phi_s}{E}{E}$ such that $$
{\mathcal
  L}_\sigma\theta=\frac{d}{ds}\Phi_s\pb\theta\at{s=0}, $$
for any
section $\theta$ of $\ext{E}$. Moreover, for every fixed $s$, the map
$\Phi_s$ is a morphism of Lie algebroids, and the base map
$\map{\phi_s}{M}{M}$, the (ordinary) flow of the vector field
$\rho(\sigma)\in\vectorfields{M}$.

\paragraph{\bf Complete lift of a section}
In this section we will define the lift of a projectable section of
$E$ to a vector field on $\Jpi$, in a similar way to the definition of
the first jet prolongations of a projectable vector field in the
standard theory of jet bundles~\cite{Saunders}.

We consider a section $\sigma$ of a Lie algebroid $E$ projectable over
a section $\eta$ of $F$. We denote by $\Psi_s$ the flow on $E$
associated to $\sigma$ and by $\Phi_s$ the flow on $F$ associated to
$\eta$. We recall that, for every fixed $s$, the maps $\Psi_s$ and
$\Phi_s$ are morphisms of Lie algebroids.  Moreover, the base maps
$\psi_s$ and $\phi_s$, are but the flows of the vector fields
$\rho_E(\sigma)$ and $\rho_F(\eta)$, respectively.

The projectability of the section implies the projectability of the
flow. It follows that (locally, in the domain of the flows) we have
defined a map $\map{\Lprol{\Psi_s}}{\Lpi}{\Lpi}$ by means of
$\Lprol{\Psi_s}(w)=\Psi_s\circ w\circ\Phi_{-s}$. By restriction of
$\Lprol{\Psi_s}$ to $\Jpi$ we get a map $\Jprol{\Psi_s}$, which is a
local flow in $\Jpi$. We will denote by $\XsigmaJ$ the vector field on
$\Jpi$ generating the flow $\Jprol{\Psi_s}$. The vector field
$\XsigmaJ$ will be called the \emph{complete lift} to $\Jpi$ of the
section $\sigma$. Since $\Jprol{\Psi_s}$ projects to the flow $\psi_s$
it follows that the vector field $\XsigmaJ$ projects to the vector
field $\rho_E(\sigma)$ in $M$.

Locally, a section $\sigma=\sigma^a e_a+\sigma^\alpha e_\alpha$ is
projectable if $\sigma^a=\sigma^a(x^i)$ depends only on $x^i$. Its
complete lift $\XsigmaJ$ has the local expression $$
\XsigmaJ=\sigma^a\pd{}{x^a}+
(\rho^A_a\sigma^a+\rho^A_\alpha\sigma^\alpha)\pd{}{u^A}+\sigma^\alpha_a\pd{}{y^\alpha_a},
$$
where $\sigma^\alpha_a =
\p{\sigma}^\alpha_{|a}+Z^\alpha_{ab}\sigma^b+Z^\alpha_{a\beta}\sigma^\beta
-y^\alpha_b\Bigl(\p{\sigma}^b_{|a}+\sigma^cC^b_{ac}\Bigr)$. In
particular, if $\sigma$ projects to the zero section, i.e.,
$\sigma^a=0$, we have $$
\XsigmaJ=\rho^A_\alpha\sigma^\alpha\pd{}{u^A}+\left(
  \p{\sigma}^\alpha_{|a}+Z^\alpha_{a\beta}\sigma^\beta
\right)\pd{}{y^\alpha_a}.  $$

\paragraph{\bf Euler-Lagrange equations} Let
$\Phi\in\Mor{\prEF}$ be a critical point of $\CMcal{S}$. In order
to find admissible variations we consider a $\prEF$-vertical
section $\sigma$ of $E$. Its flow $\map{\Psi_s}{E}{E}$ projects to
the identity in $F=TN$. Moreover we will require $\sigma$ to have
compact support. Since for every fixed $s$, the map $\Psi_s$ is a
morphism of Lie algebroids, it follows that the map
$\Phi_s=\Psi_s\circ\Phi$ is a section of $\prEF$ and a morphism of
Lie algebroids, that is, $s\mapsto\Phi_s$ is a curve in
$\Mor{\prEF}$. Using this kind of variations we have the following
result.

\begin{theorem}\cite{CFTLAVA}
  Select local coordinates such that the volume form is expressed as
  $\omega=dx^1\wedge\cdots\wedge dx^r$.  A map $\Phi$ is a critical
  section of $\CMcal{S}$ if and only if the components $y^\alpha_a$ of
  $\Phi$ satisfy the system of partial differential equations
  \begin{align*}
    &\pd{u^A}{x^a}=\rho^A_a+\rho^A_\alpha y^\alpha_a ,\\
    & \pd{y^\alpha_a}{x^b}-\pd{y^\alpha_b}{x^a}
    +C^\alpha_{b\gamma}y^\gamma_a-C^\alpha_{a\gamma}y^\gamma_b
    +C^\alpha_{\beta\gamma}y^\beta_by^\gamma_a
    +C^\alpha_{ba}=0 ,\\
    &\frac{d\,\,}{dx^a}\left(\pd{L}{y^\alpha_a}\right)
    =\pd{L}{y^\gamma_a}Z^\gamma_{a\alpha} +\pd{L}{u^A}\rho^A_\alpha.
  \end{align*}
\end{theorem}
\begin{proof}
  Recall that by $L(\Phi)$ we mean the function in $N$ given by
  $L(\Phi)(n)=L(\Phi_n)$, where $\Phi_n$ is the restriction of
  $\map{\Phi}{F}{E}$ to the fiber $F_n$. The function $L(\Phi_s)$ is
  $$
  L(\Phi_s)(n)=L(\Psi_s\circ\Phi_n)=L(\Jprol{\Psi_s}(\Phi_n))
  =(\Jprol{\Psi_s}^*L)(\Phi)(n), $$
  and therefore the variation of the
  action along the curve $s\mapsto\Phi_s$ is $$
  0=\frac{d}{ds}\CMcal{S}(\Phi_s)\at{s=0}
  =\int_N\frac{d}{ds}L(\Phi_s)\at{s=0}\omega
  =\int_N(\mathcal{L}_\XsigmaJ L)(\Phi)\,\omega.  $$
  Taking into
  account the local expression of $\XsigmaJ$ for a $\prEF$-vertical
  $\sigma$, we have that
  \begin{align*}
    \mathcal{L}_\XsigmaJ L
    &=\rho^A_\alpha\sigma^\alpha\pd{L}{u^A}+\left(
      \frac{d\sigma^\alpha}{dx^a}+Z^\alpha_{a\beta}\sigma^\beta
    \right)\pd{L}{y^\alpha_a}\\
    &=\sigma^\alpha\left[\rho^A_\alpha\pd{L}{u^A}+
      Z^\gamma_{a\alpha}\pd{L}{y^\gamma_a}
      -\frac{d}{dx^a}\left(\pd{L}{y^\alpha_a}\right)\right]
    +\frac{d}{dx^a}\left(\sigma^\alpha\pd{L}{y^\alpha_a}\right).
  \end{align*}
  Let us denote by $\delta L$ the expression with components $$
  \delta
  L_\alpha =\frac{d}{dx^a}\left(\pd{L}{y^\alpha_a}\right) -
  Z^\gamma_{a\alpha}\pd{L}{y^\gamma_a} -\rho^A_\alpha\pd{L}{u^A}, $$
  and by $J_\sigma$ the $(r-1)$-form (along $\prJN$)
  $J_\sigma=\sigma^\alpha\pd{L}{y^\alpha_a}\omega_a$ with
  $\omega_a=i_{\pd{}{x^a}}\omega$. Then we have that $$
  0=\frac{d}{ds}\CMcal{S}(\Phi_s)\at{s=0} =-\int_N(\delta
  L_\alpha\,\sigma^\alpha)\omega+\int_Nd(J_\sigma\circ\Phi).  $$
  Since
  $\sigma$ has compact support the second term vanishes by Stokes
  theorem. Moreover, since the section $\sigma$ is arbitrary, by the
  fundamental theorem of the Calculus of Variations, we get $\delta
  L=0$, which are the Euler-Lagrange equations. Notice that the first
  two equations in the above statement are nothing but the morphism
  conditions.
\end{proof}

\subsection{Examples}
\label{examples}

\subsubsection*{Standard case}
In the standard case, we consider a bundle $\map{\nu}{M}{N}$, the
standard Lie algebroids $F=TN$ and $E=TM$ and the tangent map
$\map{\prEF=T\nu}{TM}{TN}$. Then we have that $\Jpi=J^1\nu$.  When we
choose coordinate basis of vector fields (i.e., of sections of $TN$
and $TM$) we recover the equations for the standard first-order field
theory. Moreover, if we consider a different basis, what we get are
the equations for a first-order field theory written in
pseudo-coordinates~\cite{CaCrIb,Barna-L,CFTLAVA}.

\subsubsection*{Time-dependent Mechanics}
In~\cite{SaMeMa1,SaMeMa2} we developed a theory of Lagrangian
Mechanics for time dependent systems defined on Lie algebroids, where
the base manifold is fibered over the real line $\Real$.  Since
time-dependent Mechanics is nothing but a 1-dimensional field theory,
our results must reduce to that.

The morphism condition is just the admissibility condition so that, if
we write $x^0\equiv t$ and $y^\alpha_0\equiv y^\alpha$, the
Euler-Lagrange equations are $$
\frac{du^A}{dt}=\rho^A_0+\rho^A_\alpha
y^\alpha , \qquad \frac{d}{dt}\left(\pd{L}{y^\alpha}\right)
=\pd{L}{y^\gamma}(C^\gamma_{0\alpha}+C^\gamma_{\beta\alpha}y^\beta)
+\pd{L}{u^A}\rho^A_\alpha, $$
in full agreement with~\cite{MaMeSa}. In
particular, for an autonomous system on a Lie algebroid $V\to Q$, one
considers $E=T\Real\times V\to \Real\times M$ with $\prEF$ the
projection onto the first factor $T\Real$. Our results provide yet
another indication of the variational character of autonomous
mechanical systems on Lie algebroids.

\subsubsection*{Topological field theories}

Given a closed $r$-form $\Omega$ on a Lie algebroid $V\to Q$, we can
define a topological field theory as follows. For an $r$-dimensional
manifold $N$ we consider $F=TN\to N$, $E=TN\times V\to N\times Q$ and
$\prEF$ the projection onto the first factor $TN$. The Lagrangian of
the theory is $L(\Phi)=\Phi\pb\Omega$.  Then it is easy to see that
the Euler-Lagrange equations reduce to the morphism condition. In this
way, one can cope with systems such as Poisson $\sigma$-models or
Chern-Simons theories~\cite{BoKoSt, CFTLAVA, Ma6}.

\subsubsection*{Systems with symmetry}
The case of a system with symmetry is very important in Physics.  We
consider a principal bundle $\map{\nu}{P}{M}$ with structure group $G$
and we set $N=M$, $F=TN$ and $E=TP/G$ (the Atiyah algebroid of $P$),
with $\prEF([v])=T\nu(v)$. Sections of $\prEF$ are just principal
connections on $P$ and a section is a morphism if and only if it is a
flat connection. The kernel $K$ is just the adjoint bundle
$(P\times\mathfrak{g})/G\to M$.  By an adequate choice of a local
basis of sections of $F$, $K$ and $E$ one easily find the covariant
Euler-Poincar\'e equations~\cite{CaGaRa,CaRaSh}.  The covariant
Lagrange-Poincar\'e equations~\cite{CaRa} can also be recovered within
this formalism.

\section{Future work}
We have illustrated the generality of the theory of Lie algebroids and
groupoids in a wide range of situations, from nonholonomic Lagrangian
and Hamiltonian systems and mechanical control systems to Discrete
Mechanics and extensions to Field Theory. Current and future
directions of research include the following:
\begin{description}
\item[{\it Hamilton-Jacobi equation for a Hamiltonian system on a Lie
    algebroid}] It wo\-uld be interesting to continue with the study
  started in~\cite{LeMaMa} of Hamilton-Jacobi theory for Hamiltonian
  systems on Lie algebroids. In particular, it would be interesting to
  introduce a suitable definition of a local (global) complete
  integral of the Hamilton-Jacobi equation. The idea would be that the
  knowledge of an integral of the equation would allow the ``direct
  determination'' of some integral curves of the corresponding
  Hamiltonian vector field.

\item[{\it Geometric formalism for Vakonomic Mechanics on Lie
    algebroids}] An interesting topic to study is the case of
  constrained variational problems on Lie algebroids.  In this case,
  to derive the equations of motion for a Lagrangian system subject to
  nonholonomic constraints, one invokes a variational principle,
  rather than the Lagrange-D'Alembert's principle (cf.
  Section~\ref{nonlinear}).  The differential equations obtained,
  called vakonomic equations, are in general different. From an
  optimal control perspective, it seems interesting to generalize the
  formalism developed in~\cite{CLMS}.

\item[{\it Mechanical control systems on Lie algebroids}] Topics of
  interest related to mechanical control systems on Lie algebroids
  include the investigation of controllability tests along relative
  equilibria and the study of systems that include gyroscopic and
  dissipative forces.

\item[{\it Discrete Mechanics on Lie groupoids}] We are currently
  studying the construction of geometric integrators for mechanical
  systems on Lie algebroids. We have introduced the exact discrete
  Lagrangian in the Lie groupoid formalism, and are discussing
  different types of discretizations of continuous Lagrangians and
  their numerical implementation. We plan to explore natural
  extensions to forced systems and to systems with holonomic and
  nonholonomic constraints as in~\cite{CoSMa,LeMaSa}.

\item[{\it Classical Field Theory and Lie algebroids}]
  In~\cite{GrGrUr,MaMeSa} the authors have introduced the notion of a
  Lie affgebroid structure~\cite{GrGrUr1,Ma4,MaMeSa} (see also
  \cite{IMPS}).  They have developed a Lagrangian (and Hamiltonian)
  formalism on Lie affgebroids, which generalizes some classical
  formalisms for time-dependent Mechanics and, in addition, may be
  applied to other situations. Since time-dependent Mechanics is a
  $1$-dimensional field theory, it would be interesting to define the
  notion of a ``Lie multialgebroid", as a generalization of the notion
  of a Lie affgebroid. This mathematical object should encode the
  geometric structure necessary to develop field theories.  The first
  example of a Lie multialgebroid< should be ${\mathcal J}\pi$.  The
  notion of a Lie multialgebroid will potentially allow to study other
  aspects of the theory, such as Tulczyjew's triples associated with a
  Lie multialgebroid and Hamilton-Jacobi equation for classical field
  theories on Lie multialgebroids.

\end{description}

\end{document}